\documentclass[journal]{IEEEtran}
\usepackage{amsmath}
\usepackage{diagbox}
\usepackage{indentfirst}
\usepackage{graphics}
\usepackage{epsfig}
\usepackage{amsfonts}
\usepackage{amssymb}
\usepackage{subfigure}
\usepackage[numbers,square,sort&compress]{natbib}
\usepackage{caption}
\usepackage{float}
\usepackage{arydshln}
\usepackage{epstopdf}
\usepackage{color}
\usepackage{booktabs}
\usepackage{mathrsfs}
\usepackage{amsmath}
\usepackage{flushend}
\usepackage{algorithm, algorithmic}
\ifCLASSINFOpdf
\begin{document}
	\title{Observer-Based Coordinated Tracking Control for Nonlinear Multi-Agent Systems with Intermittent Communication under Heterogeneous Coupling Framework
		\thanks{This work was supported in part by  the National Natural Science Foundation of
			China under grant 61503045, in part by the Science Research Project of Education Department of Jilin Province under grant JJKH20220687KJ, and in part by the Key Science and Technology Projects of Jilin
			Province under grant 20200401075GX.
			\textit{(Corresponding author:Yulian Jiang)}}
	}
	\author{Yuhang~Zhang,~Yulian~Jiang  and~Shenquan~Wang
		\thanks{Y. Zhang,  Y. Jiang and S. Wang are with College of Electrical and Electronic
			Engineering, Changchun University of Technology, Changchun 130012, China
			(e-mail: zyh3502009@163.com; jiang\_harmony@163.com; shenquanwang@126.com).}
}
	\markboth{}%
	{Shell \MakeLowercase{\textit{et al.}}: Bare Demo of IEEEtran.cls for IEEE Transactions on Magnetics Journals}
	\maketitle
	\begin{abstract}
In this article, the observer-based coordinated tracking control problem for a class of nonlinear multi-agent systems(MASs) with intermittent communication and  information constraints is studied under dynamic switching topology.
First, a state observer is designed to estimate the unmeasurable actual state information in the system. Second, adjustable heterogeneous coupling weighting parameters are introduced in the dynamic switching topology, and the distributed coordinated tracking control protocol under heterogeneous coupling framework is proposed. Then, a new Lemma is constructed to realize the cooperative design of observer gain, state feedback gain and heterogeneous coupling gain matrices.
Furthermore, the stability of the system is further proved, and the range of communication rate is obtained.
On this basis, the intermittent communication mode is extended to three time interval cases, namely normal communication, leader-follower communication interruption and all agents communication interruption, and then the distributed coordinated tracking control method is improved to solve this problem.
Finally, simulation experiments are conducted with nonlinear MASs to verify the correctness of methods.

	\end{abstract}
	\IEEEpeerreviewmaketitle
	\section{Introduction}
	\IEEEPARstart{T}{he} problem of cooperative control of multi-agent systems (MASs) has attracted increasing attention in various scientific communities recently including synchronization of complex networks, cooperative control of multiple vehicles, coordination of multiple robotic fish, teaming of robots, and so on \cite{1,2,3,4,5,6}.
According to the control objectives of distributed control of MASs, the existing coordination control problems are divided into regulation cooperative control \cite{7} and coordinated tracking control \cite{8}.
However, it is important to design a suitable control strategy for whichever control goal you want to achieve. In the meantime, an appropriate framework should be chosen to address the phenomenon of information interaction within and outside the system.
Distributed control protocols are designed for all agents via their available local state information to achieve a specific number of interests into agreement, which can categorized into leaderless control protocol \cite{9,10} and leader-follower control protocol \cite{11,12} for MASs.
In the latter case, which is the concern of this paper, all agents are expected to follow the trajectory of expectations generated by the leader.

In recent years, many meaningful results have been generated on studies related to behavioral consensus.
State feedback based on linear combinations of relative state measurements between neighboring agents was used to implement the design of the proposed protocol in \cite{13,14,15}.
However, in practical or engineering applications, many state quantities in the system are not directly accessible because of economic costs, environmental constraints, etc., and thus the measurable output information is used to design a state observer, thus providing an elegant alternative that can be used for observation, diagnosis, and estimation purposes.

In order to estimate the leader's velocity, a new neighbor-based protocol was proposed for first-order followers in \cite{16},
and it was improved to second-order followers \cite{17,18} and high-order systems with input saturation \cite{19}, time delays \cite{20}, and external disturbance \cite{21}.
Li et al. proposed a distributed observer-based consensus protocol \cite{22} for general linear MASs and further investigated corresponding reduced-order case \cite{23}.
A fixed-time distributed observer was developed in \cite{24} with high-order integrator dynamics and matched external disturbances.
In addition, the cooperative output regulation problem of linear discrete-time time-delay MASs was developed in \cite{25} utilizing adaptive distributed observers.
Nevertheless, most of the current research on observers was focused on low-order integral systems or general linear systems. The study of observer problems in nonlinear MASs is more complicated and requires dealing with nonlinear dynamics through Lipschitz conditions.

It is noteworthy that the above distributed control problems are solved based on the fact that the communication between neighboring agents are continuous all the time.
However, communication bandwidth, physical device failures and economic costs lead to communication constraints in real systems, which means that communication between neighboring agents may be intermittent or interrupted \cite{26,27,28,29}.
How to address observer-based coordinated tracking control for nonlinear MASs with intermittent communication and dynamic switching topology via heterogeneous coupling framework is worthwhile to research, and this further motivates our work.

This article mainly studies observer-based consensus control problems for nonlinear MASs with intermittent communications under heterogeneous coupling network.
The main contributions are summarized as follows:
(1) A class of nonlinear observers is investigated based on each agent only temporally sharing its relative output with neighbors to estimate the actual and unavailable states for nonlinear MASs.
(2) The distributed tracking controllers based heterogeneous coupling framework with nonlinear observers are designed for considering intermittent constrained and intermittent coordinated constrained, in which some additional adjustable weighting parameters are constructed via dynamic interaction topology and heterogeneous coupling network.
(3) Utilizing the switching system theory, Lyapunov function and LMI technology, the stability of the system is proved and communication conditions are given, meanwhile, the co-design of heterogeneous coupling weighting, feedback gain and observer gain matrices are realized.
(4) The result presented in this work can resolve the coordinated tracking consensus for higher order nonlinear MASs with unavailable states under intermittent communication and dynamic switching topology.

	\section{Graph Theory And Problem Formulation}
\subsection{Graph Theory And Problem Formulation}

The communication topology between agents is described by
the directed graph $ G({\hat \nu } ,\zeta  ,{\tilde A})$, in which ${\hat \nu } = \left\{ {{{\hat \nu } _1},{{\hat \nu } _2},...,{{\hat \nu } _N}} \right\}$ denotes the set of nodes, $\zeta  \subseteq {\hat \nu }  \times {\hat \nu } $ denotes the set of edges, and ${\tilde A} = {\left[ {{a_{ij}}} \right]_{N \times N}}$ is the weighted adjacency matrix with elements ${a_{ij}}$. If $\left( {{{\hat \nu }_i},{{\hat \nu }_j}} \right) \in \zeta $, the information in the $jth$ node can be passed to the $ith$ one. then ${a_{ij}} = 1$. If $\left( {{{\hat \nu }_i},{{\hat \nu }_j}} \right) \notin \zeta $, ${a_{ij}} = 0$.
The directed dynamic switching topology $\tilde G$ consists of $N+1$ nodes, i.e., generated by the directed graph $G$ of the follower and the leader. Let ${L^{\vartheta \left( t \right)}}$ be the Laplace matrix of the directed graph $\tilde G$ and ${D^{\vartheta \left( t \right)}} = {\rm{diag}}\left\{ {{d_1}^{\vartheta \left( t \right)},{d_2}^{\vartheta \left( t \right)}, \ldots ,{d_N}^{\vartheta \left( t \right)}} \right\}$ denote the diagonal matrix of the different communication paths between the followers and the leader, where $\vartheta \left( t \right):\left[ {0, + \infty } \right) \to \left\{ {1,2, \ldots ,p} \right\}$ denotes the switching signal. If the followers can obtain the leader information at time $t \ge 0$, then ${d_i}^{\vartheta \left( t \right)} = 1$, otherwise ${d_i}^{\vartheta \left( t \right)} = 0$. In addition, ${\tilde G^{\vartheta \left( t \right)}} \in \bar {\tilde G } $, where $\bar {\tilde G }= \left\{ {{{\tilde G}^1}, \ldots ,{{\tilde G}^p}} \right\}$ denotes the set of topologies composed of different interaction methods under the switching signal.

\subsection{Problem Formulation}

	The nonlinear leader-following MASs considered in this research composes of $N$ agents named followers and one more agent as a leader, the dynamic model is
\begin{equation}
\begin{split}
\dot{x}_i(t)&={A}{x_i}(t)+{B}{u_i}(t)+f(x_i(t))\\
   \dot{x}_0(t)&={A}{x_0}(t)+f(x_0(t))\\
  {y_i}(t)&={C}{x_i}(t)\\
  {y_0}(t)&={C}{x_0}(t)
\end{split}
\end{equation}
where $i=1,2,...,N$, ${x_i}(t)$ $\in$ ${R}^{n}$, ${y_i}(t)$ $\in$ ${R}^z$ and ${u_i}(t)$ $\in$ ${R}^{m}$  represent the $ith$ follower's state, output and control input, respectively. In addition, ${x_0}(t)$ $\in$ ${R}^{n}$ and ${y_0}(t)$ $\in$ ${R}^z$ denote the leader's state and output. $A$, $B$ and $C$ are matrices of appropriate dimensions. $f(x_i(t))$ and $f(x_0(t))$ are continuously differentiable vector-valued functions, represent the intrinsic nonlinear dynamics of the $ith$ agent and the leader.

\emph{Assumption 1 :} The digraph $\tilde G$ contains a directed spanning tree, and the leader as one agent is the root.

\emph{Assumption 2 :}
For a nonlinear function $f\left( {{x_i}\left( t \right)} \right)$, there exists a nonnegative constant $\varpi $ such that:
\begin{equation*}
\left\| {f\left( {{x_i}\left( t \right)} \right) - f\left( {{x_0}\left( t \right)} \right)} \right\| \le \varpi \left\| {{x_i}\left( t \right) - {x_0}\left( t \right)} \right\|
\end{equation*}
then it means that $f\left( {{x_i}\left( t \right)} \right)$ satisfies the Lipschitz condition.

Some lemmas will be  provided here to derive the proposed
	criterion.
	
	\emph{Lemma 1 \cite{30}:}  Define $ L^{\vartheta \left( t \right)}$ be a structure matrix of derivative directed graph ${\tilde G}$. Then, ${\tilde G}$ has a spanning tree if and only if $ L^{\vartheta \left( t \right)}$ has a nonzero eigenvalues with multiplicity 1 and all the other nonzero eigenvalues have positive real parts.

\emph{Definition 1:} Let ${Z_n} \in {R^{n \times n}}$ be the set of all square matrices of dimension $n$ with nonpositive off-diagonal entries. A matrix $A \in {R^{n \times n}}$ is said to be a nonsingular M-matrix if $A \in {R^{n \times n}}$ and all eigenvalues of $A $ have positive real parts.

\emph{Lemma 2 \cite{31}:}  Suppose that a matrix $A = \left[ {{a_{ij}}\left( t \right)} \right] \in {R^{n \times n}}$
satisfies ${a_{ij}}\left( t \right) < 0$ for any $i \ne j$. Then, the following statements
are equivalent: 1) $A$ is a nonsingular M-matrix; 2) there exists a
positive definite $n \times n$ diagonal matrix $\Theta $ such that
${A^T}\Theta  + \Theta A > 0$; and 3) all eigenvalues of $A$ have positive real parts.

\emph{Lemma 3 \cite{32}:}  Suppose that $M \in {R^{n \times n}}$ is the symmetric positive-define matrix and $\bar R \in {R^{n \times n}}$ is the symmetric matrix, then, for any vector $x \in {R^n}$, the following formula holds
\begin{equation*}
{\lambda _{\min }}\left( {{M^{ - 1}}\bar R} \right){x^T}Mx \le {x^T}\bar Rx \le {\lambda _{\max }}\left( {{M^{ - 1}}\bar R} \right){x^T}Mx
\end{equation*}

\emph{Lemma 4 \cite{33}:}  For $\forall a,b$, one has
$
2{a^T}b \le l{a^T}a + {l^{ - 1}}{b^T}b
$,
where $\forall a,b \in {R^n}$ and $l > 0$.

\par~\\\emph{Definition 2:}
Consider the nonlinear MASs (1), if
\begin{equation}
\begin{split}
\mathop {\lim }\limits_{t \to \infty } \left\| {{x_i}\left( t \right) - {x_0}\left( t \right)} \right\| = 0 ( \forall i = 1,2, \cdots ,N)
\end{split}
\end{equation}
is satisfied for any initial states, the nonlinear MASs (1) achieve consensus tracking.

	\section{Stability Analysis}

	This session introduces the problem about observer-based tracking consensus control with intermittent communication and dynamic switching topology for nonlinear MASs, which can ensure the system (1) stable and the heterogeneous coupling gain and feedback gain matrices derived.

In order to estimate the unmeasurable states of nonlinear MASs, by using the measurable output information, a type of nonlinear observers can be designed as follows:
\begin{equation}
\begin{split}
\begin{array}{*{20}{c}}
{{{\dot {\hat x}}_i}\left( t \right) = A{{\hat x}_i}\left( t \right) + B{u_i}(t) + f\left( {{{\hat x}_i}\left( t \right)} \right)}
+ \bar G\left( {{{\hat y}_i}\left( t \right) - {y_i}\left( t \right)} \right)
\end{array}
\end{split}
\end{equation}
where ${{{\hat x}_i}\left( t \right)}$ is the estimated state of ${{x_i}\left( t \right)}$ and $\bar G$ is the observer gain matrix, $f\left( {{{\hat x}_i}\left( t \right)} \right)$ is the nonlinear function satisfying Assumption 2, ${\hat y_i}\left( t \right)$ is the output of followers.

Here, the communication mode is considered as periodic intermittent communication.
Suppose that $t \in \left[ {kw,kw + \delta } \right)=T^m$ represents normal communication time period, while ${t \in \left[ {kw + \delta ,\left( {k + 1} \right)w} \right)}=T^n$ represents communication interruption time period, where $w > \delta  > 0,k \in Z$. $t \in \left[ {kw,\left( {k + 1} \right)w} \right)$ indicates that within each periodic intermittent communication time period $w$, the normal communication time period $\delta $ and the communication interruption time period $w - \delta $ are included, where $k$ refers to the first few periodic intermittent communication periods.
In order to achieve consensus tracking for nonlinear MASs under the dynamic switching topologies with intermittent communications, the following protocol under heterogeneous coupling framework is presented:
\begin{equation}
{u_i}\left( t \right) = \left\{ {\begin{array}{*{20}{c}}
\begin{array}{l}
K\sum\limits_{j = 1}^N {{a_{ij}}^{\vartheta \left( t \right)}{\Gamma _i}^{\vartheta \left( t \right)}\left( {{{\hat x}_j}\left( t \right) - {{\hat x}_i}\left( t \right)} \right)} \\
 + K{d_i}^{\vartheta \left( t \right)}{\Gamma _i}^{\vartheta \left( t \right)}\left( {{{\hat x}_i}\left( t \right) - {x_0}\left( t \right)} \right)
\end{array}&{t \in {T^m}}\\
0&{t \in {T^n}}
\end{array}} \right.
\end{equation}
where ${A^{\vartheta \left( t \right)}} = {\left[ {{a_{ij}}^{\vartheta \left( t \right)}} \right]_{\left( {N \times N} \right)}}$ is the adjacency matrix of the directed graph ${\tilde G^{\vartheta \left( t \right)}}$, ${\Gamma _i}^{\vartheta \left( t \right)}$ represent the heterogeneous coupling gain matrix, and ${\Gamma ^{\vartheta \left( t \right)}} = diag\left\{ {{\Gamma _1}^{\vartheta \left( t \right)}, \ldots ,{\Gamma _N}^{\vartheta \left( t \right)}} \right\}$, the feedback gain matrix $K \in {R^{m \times n}}$ is designed later.

Before proposing the stable criterion,
the following errors are defined:

\begin{align}
  {e _i}\left( t \right) &= {x_i}\left( t \right) - {x_0}\left( t \right)\\
  {\Psi _i}\left( t \right) &={{\hat x}_i}\left( t \right)- {x_i}(t)
\end{align}

\begin{equation}
\begin{split}
{{\dot e}_i}\left( t \right) = \left\{ {\begin{array}{*{20}{c}}
{\begin{array}{*{20}{l}}
{A{e_i}\left( t \right) - \left( {f\left( {{x_i}\left( t \right)} \right) - f\left( {{x_0}\left( t \right)} \right)} \right)}\\
{ + BK\sum\limits_{j = 1}^N {{a_{ij}}^{\vartheta \left( t \right)}{\Gamma _i}^{\vartheta \left( t \right)}\left( {{{\hat x}_j}\left( t \right) - {{\hat x}_i}\left( t \right)} \right)} }\\
{ + BK{d_i}^{\vartheta \left( t \right)}{\Gamma _i}^{\vartheta \left( t \right)}\left( {{{\hat x}_i}\left( t \right) - {x_0}\left( t \right)} \right),t \in {T^m}}
\end{array}}\\
{A{e_i}\left( t \right) - \left( {f\left( {{x_i}\left( t \right)} \right) - f\left( {{x_0}\left( t \right)} \right)} \right),t \in {T^n}}
\end{array}} \right.
\end{split}
\end{equation}

\begin{equation}
\begin{split}
{{\dot \Psi }_i}\left( t \right) = A{\Psi _i}\left( t \right) + \left( {f\left( {{{\hat x}_i}\left( t \right)} \right) - f\left( {{x_i}\left( t \right)} \right)} \right) + GC{\Psi _i}\left( t \right)
\end{split}
\end{equation}

Define $e\left( t \right) = {\left( {{e_1}^T\left( t \right),{e_2}^T\left( t \right), \ldots ,{e_N}^T\left( t \right)} \right)^T}$ and $\Psi \left( t \right) = {\left( {{\Psi _1}^T\left( t \right),{\Psi _2}^T\left( t \right), \ldots ,{\Psi _N}^T\left( t \right)} \right)^T}$, we can get
\begin{equation}
\begin{split}
\dot e\left( t \right) = \left\{ {\begin{array}{*{20}{c}}
{\begin{array}{*{20}{l}}
{({I_N} \otimes A)e(t) - \left[ {\left( {{L^{\vartheta \left( t \right)}}{\Gamma ^{\vartheta \left( t \right)}} + {D^{\vartheta \left( t \right)}}{\Gamma ^{\vartheta \left( t \right)}}} \right)} \right.}\\
{\left. { \otimes BK} \right]\left( {e\left( t \right)\left. { + \Psi \left( t \right)} \right)} \right) + \bar f\left( {x\left( t \right)} \right),t \in {T^m}}
\end{array}}\\
{\left( {{I_N} \otimes A} \right)e\left( t \right) + \bar f\left( {x\left( t \right)} \right),t \in {T^n}}
\end{array}} \right.
\end{split}
\end{equation}

\begin{equation}
\begin{split}
\dot \Psi \left( t \right) = \left( {{I_N} \otimes A} \right)\Psi \left( t \right) + \left( {{I_N} \otimes GC} \right)\Psi \left( t \right) + \tilde f\left( {x\left( t \right)} \right)
\end{split}
\end{equation}

where $\bar f\left( {x\left( t \right)} \right) = \left( {\begin{array}{*{20}{c}}
{f\left( {{x_1}\left( t \right)} \right) - f\left( {{x_0}\left( t \right)} \right)}\\
 \vdots \\
{f\left( {{x_N}\left( t \right)} \right) - f\left( {{x_0}\left( t \right)} \right)}
\end{array}} \right)$, $\tilde f\left( {x\left( t \right)} \right) = \left( {\begin{array}{*{20}{c}}
{f\left( {{{\hat x}_1}\left( t \right)} \right) - f\left( {{x_1}\left( t \right)} \right)}\\
 \vdots \\
{f\left( {{{\hat x}_N}\left( t \right)} \right) - f\left( {{x_N}\left( t \right)} \right)}
\end{array}} \right)$.

In order to make the system achieve the tracking target (2), we give the assumption for constructing dynamic interactive network topology.

\emph{Assumption 3 :} Assume that there exists an infinite sequence of consistent bounded non-overlapping time intervals $t \in \left[ {kw,(k + 1)w} \right),k \in N$, where $\left( {k - 1} \right)w = {t_k}$, $ kw = {t_{k + 1}}$, ${t_1},{t_2}, \ldots ,{t_N}$ denotes the switching sequence of communication network topology in MASs. When $t_1=0$, ${\inf _{k \in N}}\left( {{t_{k + 1}} - {t_k}} \right) \ge {\varepsilon _0} > 0$, ${\sup _{k \in N}}\left( {{t_{k + 1}} - {t_k}} \right) < {\varepsilon _1}$.

Suppose that the switching signal is $\vartheta \left( t \right):\left[ {0, + \infty } \right) \to \left\{ {1, \ldots ,p} \right\}$. Let ${\tilde G^{\vartheta \left( t \right)}} \in \bar {\tilde G }$ be the digraph for nonlinear MASs (1), and the set of all directed graphs can be represented as $\bar{\tilde G} = \left\{ {{{\tilde G}^1}, \ldots ,{{\tilde G}^p}} \right\}$, where $p \ge 1$. After the above analysis, we are obtain, for $t \ge 0$, communication topology $\tilde G\left( t \right) = {\tilde G^{\vartheta \left( t \right)}} \in \bar {\tilde G}$, $\bar {\tilde G }= \left\{ {{{\tilde G}^1}, \ldots ,{{\tilde G}^p}} \right\}$.

Further, the following Lemma is proposed to realize the design of heterogeneous coupling gain matrix, observer gain matrix and state feedback gain matrix.

	\emph{Lemma 5 :}  Define ${\tilde G^{\vartheta \left( t \right)}},{\bar L^{\vartheta \left( t \right)}}$ and ${\Gamma ^{\vartheta \left( t \right)}}$ as the switched directed graph, its corresponding switching structure matrix and heterogeneous coupling gain matrix, respectively. Since the leader trajectory is not affected by the followers, we can obtain:
\begin{equation}
{{\bar L}^{\vartheta \left( t \right)}} = \left( {\begin{array}{*{20}{c}}
0&{0_N^T}\\
{{{\bar L}_1}^{\vartheta \left( t \right)}}&{{{\bar L}_2}^{\vartheta \left( t \right)}}
\end{array}} \right)
\end{equation}
where ${\bar L_1}^{\vartheta \left( t \right)} \in {R^N},{\bar L_2}^{\vartheta \left( t \right)} = {L^{\vartheta \left( t \right)}} + {D^{\vartheta \left( t \right)}} \in {R^{N \times N}}$, and there exists a bounded scalar $0 < \beta  \le {\lambda _{\min }}\left( {{\Lambda ^{\vartheta \left( t \right)}}} \right)$ can make the following LMIs holds
\begin{align}
{Q^1} &= \left( {\begin{array}{*{20}{c}}
\begin{array}{l}
A{P_1} + {P_1}{A^T} + {\rho ^2}{I_N}\\
 + \beta {P_1} - \beta \left( {1 - \frac{1}{2}l} \right)B{B^T}
\end{array}&{{P_1}}\\
*&{ - I}
\end{array}} \right) < 0 \\
{Q^2} &= \left( {\begin{array}{*{20}{c}}
\begin{array}{l}
A{P_2} + {P_2}{A^T}+ {\rho ^2}{I_N}\\
  + \bar Q + {{\bar Q}^T} + \beta {P_2}
\end{array}&P&{{M^T}}\\
*&{ - I}&0\\
*&*&{ - 2{\beta ^{ - 1}}l}
\end{array}} \right) < 0 \\
{Q^3} &= A{P_1} + {P_1}{A^T} + {\rho ^2}{I_N}{\rm{ + }}P_1^T{P_1} \\
{Q^4} &= A{P_2} + {P_2}{A^T} + {\rho ^2}{I_N} + \bar Q + {\bar Q^T} + P_2^T{P_2}
\end{align}
where ${\Lambda ^{\vartheta \left( t \right)}} = {\left( {{{\bar \Xi }^{\vartheta \left( t \right)}}} \right)^{ - 1}}{\Phi ^{\vartheta \left( t \right)}}{\left( {{{\bar \Xi }^{\vartheta \left( t \right)}}} \right)^{ - 1}}$,
${\bar \Xi ^{\vartheta \left( t \right)}} = \sqrt {{\Pi ^{\vartheta \left( t \right)}}{\Xi ^{\vartheta \left( t \right)}}}$,
${\hat L^{\vartheta \left( t \right)}} = \left( {{L^{\vartheta \left( t \right)}} + {D^{\vartheta \left( t \right)}}} \right){\Gamma ^{\vartheta \left( t \right)}}$,
${\Phi ^{\vartheta \left( t \right)}} = {\Pi ^{\vartheta \left( t \right)}}{\Xi ^{\vartheta \left( t \right)}}{\hat L^{\vartheta \left( t \right)}} + {\left( {{{\hat L}^{\vartheta \left( t \right)}}} \right)^T}{\Pi ^{\vartheta \left( t \right)}}{\Xi ^{\vartheta \left( t \right)}}$,
${\Xi ^{\vartheta \left( t \right)}} = diag\left\{ {{\Xi _1}^{\vartheta \left( t \right)}, \ldots ,{\Xi _p}^{\vartheta \left( t \right)}} \right\}$,
${\Xi _i}^{\vartheta \left( t \right)} = {1 \mathord{\left/
 {\vphantom {1 {{\theta _i}^{\vartheta \left( t \right)}}}} \right.
 \kern-\nulldelimiterspace} {{\theta _i}^{\vartheta \left( t \right)}}}$ and
${\theta ^{\vartheta \left( t \right)}} = {\left( {{\theta _1}^{\vartheta \left( t \right)}, \ldots ,{\theta _p}^{\vartheta \left( t \right)}} \right)^T}$, and $\left( {{L^{\vartheta \left( t \right)}} + {D^{\vartheta \left( t \right)}}} \right){\Gamma ^{\vartheta \left( t \right)}} = {I_p}$, ${\Pi ^{\vartheta \left( t \right)}} = diag\left\{ {\pi _1^{\vartheta \left( t \right)}{I_{n1}}, \ldots ,} \right.\left. {\pi _p^{\vartheta \left( t \right)}{I_{np}}} \right\}$,  $\pi _i^{\vartheta \left( t \right)}$ is the appropriate positive number to be selected. Matrices $\bar M = \bar GC{P_2}$ and $M = K{P_2}$, and feedback gain matrix $K =  - {B^T}P_1^{ - 1}$.

\begin{IEEEproof}
	It follows from Lemmas 1 and 2 that there exists a matrix ${\Xi _i}^{\vartheta \left( t \right)} > 0$ such that:
\begin{equation}
{\Xi _i}^{\vartheta \left( t \right)}\left( {L_{ii}^{\vartheta \left( t \right)} + D_{ii}^{\vartheta \left( t \right)}} \right) + {\left( {L_{ii}^{\vartheta \left( t \right)} + D_{ii}^{\vartheta \left( t \right)}} \right)^T}{\Xi _i}^{\vartheta \left( t \right)} > 0
\end{equation}

Define ${\Xi ^{\vartheta \left( t \right)}} = diag\left\{ {{\Xi _1}^{\vartheta \left( t \right)}, \ldots ,{\Xi _p}^{\vartheta \left( t \right)}} \right\}$, ${\Xi _i}^{\vartheta \left( t \right)} = {1 \mathord{\left/
 {\vphantom {1 {{\theta _i}^{\vartheta \left( t \right)}}}} \right.
 \kern-\nulldelimiterspace} {{\theta _i}^{\vartheta \left( t \right)}}}$ and ${\theta ^{\vartheta \left( t \right)}} = {\left( {{\theta _1}^{\vartheta \left( t \right)}, \ldots ,{\theta _p}^{\vartheta \left( t \right)}} \right)^T}$, and it satisfies $\left( {{L^{\vartheta \left( t \right)}} + {D^{\vartheta \left( t \right)}}} \right){\Gamma ^{\vartheta \left( t \right)}} = {I_p}$ and ${\Pi ^{\vartheta \left( t \right)}} = diag\left\{ {\pi _1^{\vartheta \left( t \right)}{I_{n1}}, \ldots ,\pi _p^{\vartheta \left( t \right)}{I_{np}}} \right\}$, next
we can obtain
\begin{equation}
{\Pi ^{\vartheta \left( t \right)}}{\Xi ^{\vartheta \left( t \right)}}\hat L_{}^{\vartheta \left( t \right)} + {\left( {\hat L_{}^{\vartheta \left( t \right)}} \right)^T}{\Pi ^{\vartheta \left( t \right)}}{\Xi ^{\vartheta \left( t \right)}} > 0
\end{equation}
Let ${\Phi _1}^{\vartheta \left( t \right)} = {\pi _1}^{\vartheta \left( t \right)}\left( {{\Xi _1}^{\vartheta \left( t \right)}{{\hat L}_{11}}^{\vartheta \left( t \right)} + {{\left( {{{\hat L}_{11}}^{\vartheta \left( t \right)}} \right)}^T}{\Xi _1}^{\vartheta \left( t \right)}} \right)$ and ${\Pi ^{\vartheta \left( t \right)}}{\Xi ^{\vartheta \left( t \right)}}{\hat L^{\vartheta \left( t \right)}}{\rm{ + }}{\left( {{{\hat L}^{\vartheta \left( t \right)}}} \right)^T}{\Pi ^{\vartheta \left( t \right)}}{\Xi ^{\vartheta \left( t \right)}} = {\Phi _p}^{\vartheta \left( t \right)}$.

It thus follows (16) that ${\Phi _1}^{\vartheta \left( t \right)} > 0$ for any positive number ${\pi _1}^{\vartheta \left( t \right)}$. Choosing a suitable ${\pi _{i + 1}}^{\vartheta \left( t \right)}$ guarantees ${\Phi _{i + 1}}^{\vartheta \left( t \right)} > 0$, in addition by Schur complement lemma and (16), we can get
\begin{align}
{\Phi _i}^{\vartheta \left( t \right)} - \pi _{i + 1}^{\vartheta \left( t \right)}{\left( {\prod\limits_{i + 1}^{\vartheta \left( t \right)} {} } \right)^T}\left( {\Xi _{i + 1}^{\vartheta \left( t \right)}\hat L_{i + 1,i + 1}^{\vartheta \left( t \right)}} \right. \nonumber \\
{\left. { + {{\left( {\hat L_{i + 1,i + 1}^{\vartheta \left( t \right)}} \right)}^T}\Xi _{i + 1}^{\vartheta \left( t \right)}} \right)^{ - 1}}\prod\limits_{i + 1}^{\vartheta \left( t \right)} {}  > 0
\end{align}
where $\prod\limits_{i + 1}^{\vartheta \left( t \right)} { = \left( {\Xi _{i + 1}^{\vartheta \left( t \right)}\hat L_{i + 1,1}^{\vartheta \left( t \right)}, \ldots ,\Xi _{i + 1}^{\vartheta \left( t \right)}\hat L_{i + 1,i}^{\vartheta \left( t \right)}} \right)} $.

It can be seen that if $\Xi _{i + 1}^{\vartheta \left( t \right)}$ is sufficiently less than $\Xi _j^{\vartheta \left( t \right)}$ for $j \le i$, equation (18) holds.  And for $i = 1, \ldots ,p$, select a group of $\pi _i^{\vartheta \left( t \right)}$, the following equation holds:
\begin{align}
{\Lambda ^{\vartheta \left( t \right)}} = {\left( {{{\bar \Xi }^{\vartheta \left( t \right)}}} \right)^{ - 1}}{\Phi ^{\vartheta \left( t \right)}}{\left( {{{\bar \Xi }^{\vartheta \left( t \right)}}} \right)^{ - 1}}
\end{align}
where ${\Lambda ^{\vartheta \left( t \right)}} > 0$ and ${\bar \Xi ^{\vartheta \left( t \right)}} = \sqrt {{\Pi ^{\vartheta \left( t \right)}}{\Xi ^{\vartheta \left( t \right)}}}$.

Assuming that $\left( {A,B} \right)$ is stable, there exists $0 < \beta  \le {\lambda _{\min }}\left( {{\Lambda ^{\vartheta \left( t \right)}}} \right)$, matrix ${P_1} > 0$, ${P_2} > 0$, ${P_3} > 0$ and ${P_4} > 0$ such that the following inequality is constructed:
\begin{align*}
A{P_1} + {P_1}{A^T} &+ {\rho ^2}{I_N} + \beta {P_1} \\
&- \beta \left( {1 - \frac{1}{2}l} \right)B{B^T} + {P_1}{P_1}^T < 0 \\
A{P_2} + {P_2}{A^T} &+ {\rho ^2}{I_N} + \bar Q + {\bar Q^T}\\
&+ \beta {P_2} + {P_2}{P_2}^T + 2\beta {l^{ - 1}}M{M^T} < 0
\end{align*}

This completes the proof. \end{IEEEproof}

Stability analysis and proofs are performed for the designed distributed coordinated tracking controllers (4) based state observers (3), and the main results are given by the following theorem.

\emph{Theorem  1:} Suppose Assumptions 1-3 hold, the nonlinear MASs (1) under switching topology with intermittent communication can achieve cooperative consensus tracking by the proposed Lemma 5, if the communication condition (20) holds:
\begin{equation}
{\delta _{\max }} > \frac{{\hat lw}}{{{{\hat \gamma }_{\min }} + \hat l}},{\delta _{\min }} > \frac{{\hat lw}}{{{{\hat \gamma }_{\max }} + \hat l}}
\end{equation}
where ${{\hat \gamma }_{\min }} = \frac{{\beta {\eta _{\min }}{\lambda _{\min }}\left( {P_1^{ - 1},P_2^{ - 1}} \right)}}{{{\eta _{\max }}{\lambda _{\max }}\left( {P_1^{ - 1},P_2^{ - 1}} \right)}}$,
${\eta _{\min \left( {\max } \right)}} = {\lambda _{\min \left( {\max } \right)}}\left( {{\Pi ^{\vartheta \left( t \right)}}{\Xi ^{\vartheta \left( t \right)}}} \right),
{{\hat \gamma }_{\max }} = \frac{{\beta {\lambda _{\min }}\left( {P_1^{ - 1},P_2^{ - 1}} \right)}}{{{\lambda _{\max }}\left( {P_1^{ - 1},P_2^{ - 1}} \right)}} , \hat l = {\lambda _{\max }}\left( {{Q^3},{Q^4}} \right){\lambda _{\max }}\left( {P_1^{ - 1},P_2^{ - 1}} \right)$,
${\lambda _{\min \left( {\max } \right)}}\left(  \cdot  \right)$  indicates the minimum (maximum) value among eigenvalues of the matrix.

	\begin{IEEEproof}
Set Lyapunov function candidate as:
\begin{align}
\nonumber V\left( t \right){\rm{ }} =& {e^T}\left( t \right)\left( {{\Pi ^{\vartheta \left( t \right)}}{\Xi ^{\vartheta \left( t \right)}} \otimes P_1^{ - 1}} \right)e\left( t \right) \\
&+ {\Psi ^T}\left( t \right)\left( {{\Pi ^{\vartheta \left( t \right)}}{\Xi ^{\vartheta \left( t \right)}} \otimes P_2^{ - 1}} \right)\Psi \left( t \right)
\end{align}
where ${\Pi ^{\vartheta \left( t \right)}}$ and ${\Xi ^{\vartheta \left( t \right)}}$ are the adjustable parameters which can be defined by Lemma 5, $P_1$ and $P_2$ are the solutions of (12) and (13). Next, the stability is analyzed in two time intervals: normal communication $t \in \left[ {kw,kw + \delta } \right)$ and communication interruption $t \in \left[ {kw + \delta ,(k + 1)w} \right)$.

For $t \in \left[ {kw,kw + \delta } \right)$
\begin{align}
\nonumber\dot V\left( t \right) =& 2{e^T}\left( t \right)\left( {{\Pi ^{\vartheta \left( t \right)}}{\Xi ^{\vartheta \left( t \right)}} \otimes P_1^{ - 1}A} \right){e^T}\left( t \right)  \\
\nonumber&+ 2{\Psi ^T}\left( t \right)\left( {{\Pi ^{\vartheta \left( t \right)}}{\Xi ^{\vartheta \left( t \right)}} \otimes P_2^{ - 1}} \right)\tilde f\left( {x\left( t \right)} \right)  \\
\nonumber &- 2{e^T}\left( t \right)\left( {\left( {{\Pi ^{\vartheta \left( t \right)}}{\Xi ^{\vartheta \left( t \right)}}} \right)\left( {{L^{\vartheta \left( t \right)}}{\Gamma ^{\vartheta \left( t \right)}}} \right.} \right.\\
\nonumber  &+ \left. {{D^{\vartheta \left( t \right)}}{\Gamma ^{\vartheta \left( t \right)}}} \right) \otimes \left. {P_1^{ - 1}BK} \right)\left( {e\left( t \right) + \Psi \left( t \right)} \right) \\
\nonumber&+ 2{e^T}\left( t \right)\left( {{\Pi ^{\vartheta \left( t \right)}}{\Xi ^{\vartheta \left( t \right)}} \otimes P_1^{ - 1}} \right)\bar f\left( {x\left( t \right)} \right)   \\
&+ 2{\Psi ^T}\left( t \right)\left( {{\Pi ^{\vartheta \left( t \right)}}{\Xi ^{\vartheta \left( t \right)}} \otimes \left( {P_2^{ - 1}A + P_2^{ - 1}GC} \right)} \right)\Psi \left( t \right)
\end{align}

Based on Lemma 4, we can obtain
\begin{align}
\nonumber&2{e^T}\left( t \right)\left( {{\Pi ^{\vartheta \left( t \right)}}{\Xi ^{\vartheta \left( t \right)}} \otimes P_1^{ - 1}} \right)\bar f\left( {x\left( t \right)} \right) \\
 &\le {e^T}\left( t \right)\left( {{\Pi ^{\vartheta \left( t \right)}}{\Xi ^{\vartheta \left( t \right)}} \otimes \left( {{\rho ^2}P_1^{ - 1}P_1^{ - 1} + I} \right)} \right){e^T}\left( t \right)
\end{align}
\begin{align}
\nonumber&
2{\Psi ^T}\left( t \right)\left( {{\Pi ^{\vartheta \left( t \right)}}{\Xi ^{\vartheta \left( t \right)}} \otimes P_2^{ - 1}} \right)\bar f\left( {x\left( t \right)} \right)\\
 & \le {\Psi ^T}\left( t \right)\left( {{\Pi ^{\vartheta \left( t \right)}}{\Xi ^{\vartheta \left( t \right)}} \otimes \left( {{\rho ^2}P_2^{ - 1}P_2^{ - 1} + I} \right)} \right){\Psi ^T}\left( t \right)
\end{align}

Substituting (23) and (24) into (22) proceeds
\begin{align}
\nonumber
\dot V\left( t \right) \le & 2{e^T}\left( t \right)\left( {{\Pi ^{\vartheta \left( t \right)}}{\Xi ^{\vartheta \left( t \right)}} \otimes P_1^{ - 1}A} \right){e^T}\left( t \right)\\
\nonumber &+ {e^T}\left( t \right)\left( {{\Pi ^{\vartheta \left( t \right)}}{\Xi ^{\vartheta \left( t \right)}} \otimes \left( {{\rho ^2}P_1^{ - 1}P_1^{ - 1} + I} \right)} \right){e^T}\left( t \right)\\
\nonumber &+ 2{\Psi ^T}\left( t \right)\left( {{\Pi ^{\vartheta \left( t \right)}}{\Xi ^{\vartheta \left( t \right)}} \otimes \left( {P_2^{ - 1}A + P_2^{ - 1}GC} \right)} \right)\Psi \left( t \right) \\
\nonumber &- 2{e^T}\left( t \right)\left( {\left( {{\Pi ^{\vartheta \left( t \right)}}{\Xi ^{\vartheta \left( t \right)}}} \right)\left( {{L^{\vartheta \left( t \right)}}{\Gamma ^{\vartheta \left( t \right)}}} \right.} \right.\\
\nonumber  &+ \left. {{D^{\vartheta \left( t \right)}}{\Gamma ^{\vartheta \left( t \right)}}} \right) \otimes \left. {P_1^{ - 1}BK} \right)\left( {e\left( t \right) + \Psi \left( t \right)} \right) \\
&+ {\Psi ^T}\left( t \right)\left( {{\Pi ^{\vartheta \left( t \right)}}{\Xi ^{\vartheta \left( t \right)}} \otimes \left( {{\rho ^2}P_2^{ - 1}P_2^{ - 1} + I} \right)} \right){\Psi ^T}\left( t \right)
\end{align}

According to Lemma 3, we have
\begin{align}
\nonumber&
2{e^T}\left( t \right)\left( {\left( {{\Pi ^{\vartheta \left( t \right)}}{\Xi ^{\vartheta \left( t \right)}}{L^{\vartheta \left( t \right)}}{\Gamma ^{\vartheta \left( t \right)}}} \right)} \right. \otimes \left( {P_1^{ - 1}BK} \right)\Psi \left( t \right) \\
\nonumber & + 2{e^T}\left( t \right)\left( {\left( {{\Pi ^{\vartheta \left( t \right)}}{\Xi ^{\vartheta \left( t \right)}}{D^{\vartheta \left( t \right)}}{\Gamma ^{\vartheta \left( t \right)}}} \right)} \right. \otimes \left( {P_1^{ - 1}BK} \right)\Psi \left( t \right) \\
\nonumber   \le &  {e^T}\left( t \right)\left[ {\left( {{\Pi ^{\vartheta \left( t \right)}}{\Xi ^{\vartheta \left( t \right)}}{{\hat L}^{\vartheta \left( t \right)}}} \right) \otimes \left( {lP_1^{ - 1}B} \right){{\left( {P_1^{ - 1}B} \right)}^T}} \right]e\left( t \right)\\
&+  {\Psi ^T}\left( t \right)\left[ {\left( {{\Pi ^{\vartheta \left( t \right)}}{\Xi ^{\vartheta \left( t \right)}}{{\hat L}^{\vartheta \left( t \right)}}} \right) \otimes \left( {{l^{ - 1}}{F^T}F} \right)} \right]\Psi \left( t \right)
\end{align}

Then, substituting $K =  - {B^T}{P_1}^{ - 1}$ into (25) yields
\begin{align}
\nonumber
\dot V\left( t \right) \le & {e^T}\left( t \right)\left( {{\Pi ^{\vartheta \left( t \right)}}{\Xi ^{\vartheta \left( t \right)}} \otimes \left( {P_1^{ - 1}A + {A^T}P_1^{ - 1}} \right)} \right)e\left( t \right)\\
\nonumber &+ {e^T}\left( t \right)\left( {{\Pi ^{\vartheta \left( t \right)}}{\Xi ^{\vartheta \left( t \right)}} \otimes \left( {{\rho ^2}P_1^{ - 1}P_1^{ - 1} + I} \right)} \right)e\left( t \right)\\
\nonumber & + {\Psi ^T}\left( t \right)\left( {{\Pi ^{\vartheta \left( t \right)}}{\Xi ^{\vartheta \left( t \right)}} \otimes \left( {P_2^{ - 1}A + {A^T}P_2^{ - 1}} \right)} \right)\Psi \left( t \right)\\
\nonumber & + {\Psi ^T}\left( t \right)\left( {{\Pi ^{\vartheta \left( t \right)}}{\Xi ^{\vartheta \left( t \right)}} \otimes \left( {2P_2^{ - 1}GC} \right)} \right)\Psi \left( t \right)\\
\nonumber &  + 2{e^T}\left( t \right)\left( {\left( {{\Pi ^{\vartheta \left( t \right)}}{\Xi ^{\vartheta \left( t \right)}}{{\hat L}^{\vartheta \left( t \right)}}} \right) \otimes \hat P} \right)e\left( t \right) \\
\nonumber &+ 2{e^T}\left( t \right)\left( {\left( {{{\left( {{{\hat L}^{\vartheta \left( t \right)}}} \right)}^T}{\Pi ^{\vartheta \left( t \right)}}{\Xi ^{\vartheta \left( t \right)}}} \right) \otimes \hat P} \right)e\left( t \right) \\
\nonumber &+ {\Psi ^T}\left( t \right)\left( {{\Pi ^{\vartheta \left( t \right)}}{\Xi ^{\vartheta \left( t \right)}} \otimes \left( {{\rho ^2}P_2^{ - 1}P_2^{ - 1} + I} \right)} \right){\Psi ^T}\left( t \right)\\
\nonumber &- {e^T}\left( t \right)\left[ {\left( {{\Pi ^{\vartheta \left( t \right)}}{\Xi ^{\vartheta \left( t \right)}}{{\hat L}^{\vartheta \left( t \right)}}} \right) \otimes l\hat P} \right]e\left( t \right)\\
 &- {\Psi ^T}\left( t \right)\left[ {\left( {{\Pi ^{\vartheta \left( t \right)}}{\Xi ^{\vartheta \left( t \right)}}{{\hat L}^{\vartheta \left( t \right)}}} \right) \otimes \left( {{l^{ - 1}}\hat P} \right)} \right]\Psi \left( t \right)
\end{align}
where $\hat P=P_1^{ - 1}B{B^T}P_1^{ - 1}$, and define $\mathord{\buildrel{\lower3pt\hbox{$\scriptscriptstyle\frown$}}
\over \varepsilon } \left( t \right) = {\left( {{{\mathord{\buildrel{\lower3pt\hbox{$\scriptscriptstyle\frown$}}
\over \varepsilon } }_1}^T\left( t \right), \ldots ,{{\mathord{\buildrel{\lower3pt\hbox{$\scriptscriptstyle\frown$}}
\over \varepsilon } }_N}^T\left( t \right)} \right)^T}$, $\mathord{\buildrel{\lower3pt\hbox{$\scriptscriptstyle\smile$}}
\over \Psi } \left( t \right) = {\left( {{{\mathord{\buildrel{\lower3pt\hbox{$\scriptscriptstyle\smile$}}
\over \Psi } }_1}^T\left( t \right), \ldots ,{{\mathord{\buildrel{\lower3pt\hbox{$\scriptscriptstyle\smile$}}
\over \Psi } }_N}^T\left( t \right)} \right)^T}$, ${\mathord{\buildrel{\lower3pt\hbox{$\scriptscriptstyle\frown$}}
\over \varepsilon } _i}\left( t \right) = \sqrt {{\Pi ^{\vartheta \left( t \right)}}{\Xi ^{\vartheta \left( t \right)}}} {P_1}^{ - 1}{e_i}\left( t \right)$, ${\mathord{\buildrel{\lower3pt\hbox{$\scriptscriptstyle\smile$}}
\over \Psi } _i}\left( t \right) = \sqrt {{\Pi ^{\vartheta \left( t \right)}}{\Xi ^{\vartheta \left( t \right)}}} {P_2}^{ - 1}{\Psi _i}\left( t \right)$.
Then, we have $e\left( t \right) = \left( {{{\left( {\sqrt {{\Pi ^{\vartheta \left( t \right)}}{\Xi ^{\vartheta \left( t \right)}}} } \right)}^{ - 1}} \otimes {P^{ - 1}}} \right){\mathord{\buildrel{\lower3pt\hbox{$\scriptscriptstyle\frown$}}
\over \varepsilon } ^T}\left( t \right)$,  $\Psi \left( t \right) = \left( {{{\left( {\sqrt {{\Pi ^{\vartheta \left( t \right)}}{\Xi ^{\vartheta \left( t \right)}}} } \right)}^{ - 1}} \otimes {P_2}^{ - 1}} \right){\mathord{\buildrel{\lower3pt\hbox{$\scriptscriptstyle\smile$}}
\over \Psi } ^T}\left( t \right)$.

(27) can be written as

\begin{align}
\nonumber\dot V\left( t \right) \le & {{\mathord{\buildrel{\lower3pt\hbox{$\scriptscriptstyle\frown$}}
\over \varepsilon } }^T}\left( t \right)\left( {{I_N} \otimes \left( {A{P_1} + {P_1}{A^T} + {\rho ^2} + {P_1}^T{P_1}} \right)} \right)\mathord{\buildrel{\lower3pt\hbox{$\scriptscriptstyle\frown$}}
\over \varepsilon } \left( t \right)\\
\nonumber &- {{\mathord{\buildrel{\lower3pt\hbox{$\scriptscriptstyle\frown$}}
\over \varepsilon } }^T}\left( t \right)\left( {{\Lambda ^{\vartheta \left( t \right)}} \otimes \left( {B{B^T} + \frac{1}{2}lB{B^T}} \right)} \right)\mathord{\buildrel{\lower3pt\hbox{$\scriptscriptstyle\frown$}}
\over \varepsilon } \left( t \right)\\
\nonumber &+ {{\mathord{\buildrel{\lower3pt\hbox{$\scriptscriptstyle\smile$}}
\over \Psi } }^T}\left( t \right)\left( {{I_N} \otimes \left( {A{P_2} + {P_2}{A^T}} \right)} \right)\mathord{\buildrel{\lower3pt\hbox{$\scriptscriptstyle\smile$}}
\over \Psi } \left( t \right)\\
\nonumber &+ {{\mathord{\buildrel{\lower3pt\hbox{$\scriptscriptstyle\smile$}}
\over \Psi } }^T}\left( t \right)\left( {{I_N} \otimes \left( {2GC{P_2} + {\rho ^2} + P_2^T{P_2}} \right)} \right)\mathord{\buildrel{\lower3pt\hbox{$\scriptscriptstyle\smile$}}
\over \Psi } \left( t \right) \\
 &- \frac{1}{2}{{\mathord{\buildrel{\lower3pt\hbox{$\scriptscriptstyle\smile$}}
\over \Psi } }^T}\left( t \right)\left[ {{\Lambda ^{\vartheta \left( t \right)}} \otimes \left( {{l^{ - 1}}F{P_2}{{\left( {F{P_2}} \right)}^T}} \right)} \right]\mathord{\buildrel{\lower3pt\hbox{$\scriptscriptstyle\smile$}}
\over \Psi } \left( t \right)
\end{align}

According to Lemmas 2, 3 and 5, we can obtain $0 < \beta  \le {\lambda _{\min }}\left( {{\Lambda ^{\vartheta \left( t \right)}}} \right)$, and the time derivative of $V(t)$ for the communication network topology ${\tilde G^{\vartheta \left( t \right)}}$ under any switching signal $\vartheta \left( t \right)$  is taken as:
\begin{align}
\nonumber \dot V\left( t \right) \le & {{\mathord{\buildrel{\lower3pt\hbox{$\scriptscriptstyle\frown$}}
\over \varepsilon } }^T}\left( t \right)\left( {{I_N} \otimes \left( {A{P_1} + {P_1}{A^T} + {\rho ^2} + {P_1}^T{P_1}} \right)} \right)\mathord{\buildrel{\lower3pt\hbox{$\scriptscriptstyle\frown$}}
\over \varepsilon } \left( t \right)\\
\nonumber &- {{\mathord{\buildrel{\lower3pt\hbox{$\scriptscriptstyle\frown$}}
\over \varepsilon } }^T}\left( t \right)\left( {{I_N} \otimes \left( {\beta \left( {B{B^T} + \frac{1}{2}lB{B^T}} \right)} \right)} \right)\mathord{\buildrel{\lower3pt\hbox{$\scriptscriptstyle\frown$}}
\over \varepsilon } \left( t \right)\\
\nonumber &+ {{\mathord{\buildrel{\lower3pt\hbox{$\scriptscriptstyle\smile$}}
\over \Psi } }^T}\left( t \right)\left( {{I_N} \otimes \left( {A{P_2} + {P_2}{A^T} + 2GC{P_2}} \right)} \right)\mathord{\buildrel{\lower3pt\hbox{$\scriptscriptstyle\smile$}}
\over \Psi } \left( t \right)\\
\nonumber &+ {{\mathord{\buildrel{\lower3pt\hbox{$\scriptscriptstyle\smile$}}
\over \Psi } }^T}\left( t \right)\left( {{I_N} \otimes \left( {{\rho ^2} + P_2^T{P_2}} \right)} \right)\mathord{\buildrel{\lower3pt\hbox{$\scriptscriptstyle\smile$}}
\over \Psi } \left( t \right)\\
 &- {{\mathord{\buildrel{\lower3pt\hbox{$\scriptscriptstyle\smile$}}
\over \Psi } }^T}\left( t \right)\left( {{I_N} \otimes \left( {\frac{1}{2}\beta {l^{ - 1}}F{P_2}{{\left( {F{P_2}} \right)}^T}} \right)} \right)\mathord{\buildrel{\lower3pt\hbox{$\scriptscriptstyle\smile$}}
\over \Psi } \left( t \right)
\end{align}

Combining (12) with (13) in Lemma 5, (29) can be written as
\begin{align}
\nonumber \dot V\left( t \right)  \le & - \beta {e^T}\left( t \right)\left( {{\Pi ^{\vartheta \left( t \right)}}{\Xi ^{\vartheta \left( t \right)}} \otimes P_1^{ - 1}} \right)e\left( t \right)\\
 &- \beta {\Psi ^T}\left( t \right)\left( {{\Pi ^{\vartheta \left( t \right)}}{\Xi ^{\vartheta \left( t \right)}} \otimes P_2^{ - 1}} \right)\Psi \left( t \right)
\end{align}

To obtain the maximum value of ${\delta _{\max }}$, define ${\eta _{\max }} = {\lambda _{\max }}\left( {{\Pi ^{\vartheta \left( t \right)}}{\Xi ^{\vartheta \left( t \right)}}} \right)$, ${\eta _{\min }} = {\lambda _{\min }}\left( {{\Pi ^{\vartheta \left( t \right)}}{\Xi ^{\vartheta \left( t \right)}}} \right)$. According to (19) and Lemma 3, substituting $V\left( t \right) \le {\lambda _{\max }}\left( {P_1^{ - 1},P_2^{ - 1}} \right){\eta _{\max }}{e^T}\left( t \right)e\left( t \right) + {\rm{ + }}{\lambda _{\max }}\left( {P_1^{ - 1},P_2^{ - 1}} \right){\eta _{\max }}{\Psi ^T}\left( t \right)\Psi \left( t \right)$ into ( )

\begin{align}
\nonumber \dot V\left( t \right) \le & - \beta {\lambda _{\min }}{\left( {P_1^{ - 1},P_2^{ - 1}} \right)^T}e\left( t \right)\left( {{\Pi ^{\vartheta \left( t \right)}}{\Xi ^{\vartheta \left( t \right)}} \otimes {I_N}} \right)e\left( t \right)\\
\nonumber  & - \beta {\lambda _{\min }}{\left( {P_2^{ - 1},P_2^{ - 1}} \right)^T}\Psi \left( t \right)\left( {{\Pi ^{\vartheta \left( t \right)}}{\Xi ^{\vartheta \left( t \right)}} \otimes {I_N}} \right)\Psi \left( t \right)\\
 \le & - \hat \gamma V\left( t \right)
\end{align}
where ${{\hat \gamma }_{\min }} = \frac{{\beta {\eta _{\min }}{\lambda _{\min }}\left( {P_1^{ - 1},P_2^{ - 1}} \right)}}{{{\eta _{\max }}{\lambda _{\max }}\left( {P_1^{ - 1},P_2^{ - 1}} \right)}}$, $P_1$ and $P_2$ are the positive-definition matrix, so we can get $\hat \gamma  > 0$. Similar to (31), by defining ${{\hat \gamma }_{\max }} = \frac{{\beta {\lambda _{\min }}\left( {P_1^{ - 1},P_2^{ - 1}} \right)}}{{{\lambda _{\max }}\left( {P_1^{ - 1},P_2^{ - 1}} \right)}}$, and according to $V\left( t \right) \le {\lambda _{\max }}\left( {P_1^{ - 1},P_2^{ - 1}} \right){e^T}\left( t \right)\left( {{\Pi ^{\vartheta \left( t \right)}}{\Xi ^{\vartheta \left( t \right)}} \otimes {I_N}} \right)e\left( t \right) + {\rm{ + }}{\lambda _{\max }}\left( {P_1^{ - 1},P_2^{ - 1}} \right){\Psi ^T}\left( t \right)\left( {{\Pi ^{\vartheta \left( t \right)}}{\Xi ^{\vartheta \left( t \right)}} \otimes {I_N}} \right)\Psi \left( t \right)$ to obtain the minimum value of ${\delta _{\min }}$.

For $t \in \left[ {kw + \delta ,(k + 1)w} \right)$,

\begin{align}
\nonumber
\dot V\left( t \right) =& 2{e^T}\left( t \right)\left( {{\Pi ^{\vartheta \left( t \right)}}{\Xi ^{\vartheta \left( t \right)}} \otimes P_1^{ - 1}A} \right){e^T}\left( t \right)\\
\nonumber &+ 2{\Psi ^T}\left( t \right)\left( {{\Pi ^{\vartheta \left( t \right)}}{\Xi ^{\vartheta \left( t \right)}} \otimes P_2^{ - 1}} \right)\tilde f\left( {x\left( t \right)} \right)\\
\nonumber &+ 2{e^T}\left( t \right)\left( {{\Pi ^{\vartheta \left( t \right)}}{\Xi ^{\vartheta \left( t \right)}} \otimes P_1^{ - 1}} \right)\bar f\left( {x\left( t \right)} \right)\\
 & + 2{\Psi ^T}\left( t \right)\left( {{\Pi ^{\vartheta \left( t \right)}}{\Xi ^{\vartheta \left( t \right)}} \otimes \left( {P_2^{ - 1}A + P_2^{ - 1}GC} \right)} \right)\Psi \left( t \right)
\end{align}

Similarly, based on Lemmas 3 and 4, and some transformations of the necessary form, equation (32) is rewritten as
\begin{align}
\nonumber
\dot V\left( t \right) \le & {{\mathord{\buildrel{\lower3pt\hbox{$\scriptscriptstyle\frown$}}
\over \varepsilon } }^T}\left( t \right)\left( {{I_N} \otimes \left( {A{P_1} + {P_1}{A^T}} \right)} \right)\mathord{\buildrel{\lower3pt\hbox{$\scriptscriptstyle\frown$}}
\over \varepsilon } \left( t \right)\\
\nonumber  &+ {{\mathord{\buildrel{\lower3pt\hbox{$\scriptscriptstyle\frown$}}
\over \varepsilon } }^T}\left( t \right)\left( {{I_N} \otimes \left( {{\rho ^2} + {P_1}^T{P_1}} \right)} \right)\mathord{\buildrel{\lower3pt\hbox{$\scriptscriptstyle\frown$}}
\over \varepsilon } \left( t \right)\\
\nonumber  & + {{\mathord{\buildrel{\lower3pt\hbox{$\scriptscriptstyle\smile$}}
\over \Psi } }^T}\left( t \right)\left( {{I_N} \otimes \left( {A{P_2} + {P_2}{A^T} + {\rho ^2}} \right)} \right)\mathord{\buildrel{\lower3pt\hbox{$\scriptscriptstyle\smile$}}
\over \Psi } \left( t \right)\\
 & + {{\mathord{\buildrel{\lower3pt\hbox{$\scriptscriptstyle\smile$}}
\over \Psi } }^T}\left( t \right)\left( {{I_N} \otimes \left( {2GC{P_2} + P_2^T{P_2}} \right)} \right)\mathord{\buildrel{\lower3pt\hbox{$\scriptscriptstyle\smile$}}
\over \Psi } \left( t \right)
\end{align}

Utilizing Lemma 5 and
$V\left( t \right) \le {\lambda _{\max }}\left( {P_1^{ - 1},P_2^{ - 1}} \right){e^T}\left( t \right)\left( {{\Pi ^{\vartheta \left( t \right)}}{\Xi ^{\vartheta \left( t \right)}} \otimes {I_N}} \right)e\left( t \right) + {\lambda _{\max }}\left( {P_1^{ - 1},P_2^{ - 1}} \right){\Psi ^T}\left( t \right)\left( {{\Pi ^{\vartheta \left( t \right)}}{\Xi ^{\vartheta \left( t \right)}} \otimes {I_N}} \right)\Psi \left( t \right)$, (31) can be written as
\begin{align}
\nonumber \dot V\left( t \right) \le & {\lambda _{\max }}{\left( {{Q^3},{Q^4}} \right)^T}{{\mathord{\buildrel{\lower3pt\hbox{$\scriptscriptstyle\frown$}}
\over \varepsilon } }^T}\left( t \right)\mathord{\buildrel{\lower3pt\hbox{$\scriptscriptstyle\frown$}}
\over \varepsilon } \left( t \right)\\
\nonumber & + {\lambda _{\max }}{\left( {{Q^3},{Q^4}} \right)^T}{{\mathord{\buildrel{\lower3pt\hbox{$\scriptscriptstyle\smile$}}
\over \Psi } }^T}\left( t \right)\mathord{\buildrel{\lower3pt\hbox{$\scriptscriptstyle\smile$}}
\over \Psi } \left( t \right)\\
 \le & \hat lV\left( t \right)
\end{align}
where $\hat l = {\lambda _{\max }}\left( {{Q^3},{Q^4}} \right){\lambda _{\max }}\left( {P_1^{ - 1},P_2^{ - 1}} \right)$, $P_1$ and $P_2$ are the positive-definition matrix, so we can get $\hat l > 0$. According to the minimum value ${\hat \gamma _{\min }}$ and the maximum value ${\hat \gamma _{\max }}$, we can obtain

\begin{align}
V\left( w \right) \le V\left( {{\delta _0}} \right){e^{\hat l\left( {w - {\delta _0}} \right)}} \le V\left( 0 \right){e^{ - \hat \gamma {\delta _0} + \hat l\left( {w - {\delta _0}} \right)}} = V\left( 0 \right){e^{ - {{\tilde \Upsilon }_1}}}
\end{align}
where ${\tilde \Upsilon _1}{ = ^{ - \hat \gamma {\delta _0} + \hat l\left( {w - {\delta _0}} \right)}}$. The range of ${\tilde \Upsilon _1}$ can be obtained by (20), that is ${\tilde \Upsilon _1} > 0$. For any positive integer $k$,

\begin{align}
V\left( {\left( {k + 1} \right)w} \right) \le V\left( 0 \right){e^{ - \sum\limits_{j = 1}^k {{{\tilde \Upsilon }_j}} }}
\end{align}
where ${\tilde \Upsilon _j} = \hat \gamma {\delta _j} - \hat l\left( {w - {\delta _j}} \right) > 0$, $j = 1, \ldots ,k$. Set $sw \le t \le \left( {s + 1} \right)w$ for any $s \in R $. Define $\left[ {kw,(k + 1)w} \right)$, $k \in {{\rm{N}}_ + }$  is a uniformly bounded time sequence, ${\mathord{\buildrel{\lower3pt\hbox{$\scriptscriptstyle\smile$}}
\over \omega } _{\max }} = {\max _{k \in {_ + }}}\left\{ {\left( {k + 1} \right)w,kw} \right\}$ and $\tilde \Upsilon  = {\min _{j \in {{\rm{N}}_ + }}}\left\{ {{{\tilde \Upsilon }_j}} \right\}$. Finally, according to the recursion principle, we can obtain
\begin{align}
\nonumber
V\left( t \right) & \le V\left( {kw} \right){e^{{{\mathord{\buildrel{\lower3pt\hbox{$\scriptscriptstyle\smile$}}
\over \omega } }_{\max }}\hat l}}\\
\nonumber & \le {e^{{{\mathord{\buildrel{\lower3pt\hbox{$\scriptscriptstyle\smile$}}
\over \omega } }_{\max }}\hat l}}V\left( 0 \right){e^{ - \sum\nolimits_{j = 1}^s {{{\tilde \Upsilon }_j}} }}\\
\nonumber & \le {e^{{{\mathord{\buildrel{\lower3pt\hbox{$\scriptscriptstyle\smile$}}
\over \omega } }_{\max }}\hat l}}V\left( 0 \right){e^{ - \frac{{\tilde \Upsilon }}{{{{\mathord{\buildrel{\lower3pt\hbox{$\scriptscriptstyle\smile$}}
\over \omega } }_{\max }}}}t}}\\
& = {\Omega _0}{e^{ - {\Omega _1}t}}
\end{align}
where ${\Omega _0} = {e^{{{\mathord{\buildrel{\lower3pt\hbox{$\scriptscriptstyle\smile$}}
\over \omega } }_{\max }}\hat l}}{V}\left( 0 \right)$ and ${\Omega _1} = {{\tilde \Upsilon } \mathord{\left/
 {\vphantom {{\tilde \Upsilon } {{{\mathord{\buildrel{\lower3pt\hbox{$\scriptscriptstyle\smile$}}
\over \omega } }_{\max }}}}} \right.
 \kern-\nulldelimiterspace} {{{\mathord{\buildrel{\lower3pt\hbox{$\scriptscriptstyle\smile$}}
\over \omega } }_{\max }}}}$. Through the above process, it is demonstrated that the nonlinear MASa (1) under a dynamically switched communication network topology subject to the combined constraints of intermittent communication and information unpredictability is able to achieve the consensus tracking goal (2) under the combined action of a state observer (3) and a distributed coordinated tracking controller (4).

This completes the proof. \end{IEEEproof}

Theorem 1 and Lemma 5 mainly address the distributed coordinated tracking control problem for nonlinear MASs when intermittent communication is divided into normal communication $t \in \left[ {kw,kw + \delta } \right)$ and communication interruption $t \in \left[ {kw + \delta ,\left( {k + 1} \right)w} \right)$. However, in some specific leader-follower MASs, communication interruption is extended to leader-follower communication interruption $t \in \left[ {kw + \delta ,kw + h} \right)$ and all agents communication interruption $t \in \left[ {kw + h,\left( {k + 1} \right)w} \right)$.
When the intermittent communication method is extended as described above, the distributed coordinated tracking control protocol can be further improved and extended:
\begin{equation}
{u_i}\left( t \right) = \left\{ {\begin{array}{*{20}{c}}
\begin{array}{l}
{K'}\sum\limits_{j = 1}^N {{a_{ij}}^{\vartheta \left( t \right)}{\Gamma _i'}^{\vartheta \left( t \right)}\left( {{{\hat x}_j}\left( t \right) - {{\hat x}_i}\left( t \right)} \right)} \\
 + {K'}{d_i}^{\vartheta \left( t \right)}{\Gamma _i'}^{\vartheta \left( t \right)}\left( {{{\hat x}_i}\left( t \right) - {x_0}\left( t \right)} \right),t \in {T^m}
\end{array}\\
{{K'}\sum\limits_{j = 1}^N {{a_{ij}}^{\vartheta \left( t \right)}{\Gamma _i'}^{\vartheta \left( t \right)}\left( {{{\hat x}_j}\left( t \right) - {{\hat x}_i}\left( t \right)} \right),t \in {T^q}} }\\
{0,t \in {T^n}}
\end{array}} \right.
\end{equation}
where $t \in \left[ {kw,kw + \delta } \right)=T^m$,
$\left[ {kw + \delta ,kw + h} \right)={T^q} $ and ${t \in \left[ {kw + h ,\left( {k + 1} \right)w} \right)}=T^n$.

In order to design the heterogeneous coupling gain matrix ${\Gamma _i}^{\vartheta \left( t \right)}$ and state feedback gain matrix ${K'} \in {R^{m \times n}}$ in the distributed coordinated tracking control protocol (38) and achieve the tracking target (2), we construct the following Lemma and Corollary.

\emph{Lemma 6 :}
Similar to Lemma 5, there exists a bounded positive scalar $\beta$, and the switching consensus protocol (38) under
heterogeneous coupling framework with intermittent communication and dynamic switching topology for nonlinear MASs (1)
can be constructed by following matrices:

\begin{align}
{M^1} &= \left( {\begin{array}{*{20}{c}}
\begin{array}{l}
A{P_1'} + {P_1'}{A^T} + {\rho ^2}{I_N}\\
 + \beta {P_1'} - \beta \left( {1 - \frac{1}{2}l} \right)B{B^T}
\end{array}&{{P_1'}}\\
*&{ - I}
\end{array}} \right) < 0 \\
{M^2} &= \left( {\begin{array}{*{20}{c}}
\begin{array}{l}
A{P_2'} + {P_2'}{A^T}+ {\rho ^2}{I_N}\\
  + {\bar Q}' + {{\bar Q}'^T} + \beta {P_2'}
\end{array}&P_2'&{{M^T}}\\
*&{ - I}&0\\
*&*&{ - 2{\beta ^{ - 1}}l}
\end{array}} \right) < 0 \\
{M^3} &=A{P_1'} + {P_1'}{A^T} + {\rho ^2}{I_N}+{P_1'}^T{P_1'}
 - \beta \left( {1 - \frac{1}{2}l} \right)B{B^T}\\
{M^4} &=A{P_2'} + {P_2'}{A^T}+ {\rho ^2}{I_N}
  + {\bar Q}' + {{\bar Q}'^T}+{P_2'}^T{P_2'}\\
{M^5} &= A{P_1'} + {P_1'}{A^T} + {\rho ^2}{I_N}{\rm{ + }}P_1'^T{P_1'} \\
{M^6} &= A{P_2'} + {P_2'}{A^T} + {\rho ^2}{I_N} + {\bar Q}' + {{\bar Q}'^T} + P_2'^T{P_2'}
\end{align}

\emph{Corollary 1:} Suppose Assumptions 1-3 hold, the nonlinear HMASs (1) under switching topology with intermittent communications can achieve cooperative consensus tracking by the proposed Lemma 6, if the following condition (45) holds:
\begin{align}
\nonumber & \frac{{{{\hat \gamma' }_{\max }}{\delta _j} + \hat l'\left( {{h_j} + {\delta _j}} \right) - \mathord{\buildrel{\lower3pt\hbox{$\scriptscriptstyle\frown$}}
\over \chi } \left( {{h_j} - {\delta _j}} \right)}}{{\hat l'}}\\
 \ge &  \frac{{{{\hat \gamma' }_{\min }}{\delta _j} + \hat l'\left( {{h_j} + {\delta _j}} \right) - \mathord{\buildrel{\lower3pt\hbox{$\scriptscriptstyle\frown$}}
\over \chi } \left( {{h_j} - {\delta _j}} \right)}}{{\hat l'}} > \hat t
\end{align}
where ${{\hat \gamma' }_{\min }} = \frac{{\beta {\eta _{\min }}{\lambda _{\min }}\left( {P_1'^{ - 1},P_2'^{ - 1}} \right)}}{{{\eta _{\max }}{\lambda _{\max }}\left( {P_1'^{ - 1},P_2'^{ - 1}} \right)}}$,
${\eta _{\min \left( {\max } \right)}} = {\lambda _{\min \left( {\max } \right)}}\left( {{\Pi ^{\vartheta \left( t \right)}}{\Xi ^{\vartheta \left( t \right)}}} \right),
{{\hat \gamma'}_{\max }} = \frac{{\beta {\lambda _{\min }}\left( {P_1'^{ - 1},P_2'^{ - 1}} \right)}}{{{\lambda _{\max }}\left( {P_1'^{ - 1},P_2'^{ - 1}} \right)}} , \hat l = {\lambda _{\max }}\left( {{M^3},{M^4}} \right){\lambda _{\max }}\left( {P_1'^{ - 1},P_2'^{ - 1}} \right)$, $\mathord{\buildrel{\lower3pt\hbox{$\scriptscriptstyle\frown$}}
\over \chi }  = {\lambda _{\max }}\left( {{M^5},{M^6}} \right){\lambda _{\max }}\left( {P_1'^{ - 1},P_2'^{ - 1}} \right)$,
${\lambda _{\min \left( {\max } \right)}}\left(  \cdot  \right)$  indicates the minimum (maximum) value among eigenvalues of the matrix.

	\begin{IEEEproof}
Set Lyapunov function candidate as:
\begin{align}
\nonumber V'\left( t \right){\rm{ }} =& {e^T}\left( t \right)\left( {{\Pi ^{\vartheta \left( t \right)}}{\Xi ^{\vartheta \left( t \right)}} \otimes P_1'^{ - 1}} \right)e\left( t \right) \\
&+ {\Psi ^T}\left( t \right)\left( {{\Pi ^{\vartheta \left( t \right)}}{\Xi ^{\vartheta \left( t \right)}} \otimes P_2'^{ - 1}} \right)\Psi \left( t \right)
\end{align}
where ${\Pi ^{\vartheta \left( t \right)}}$ and ${\Xi ^{\vartheta \left( t \right)}}$ are the adjustable parameters which can be defined by Lemma 6, $P_1'$ and $P_2'$ are the solutions of (39) and (40).

When $t \in \left[ {kw,kw + \delta } \right)$, by the similar process in Theorem 1, we can obtain
\begin{align}
\nonumber\dot V'\left( t \right) =& 2{e^T}\left( t \right)\left( {{\Pi ^{\vartheta \left( t \right)}}{\Xi ^{\vartheta \left( t \right)}} \otimes P_1'^{ - 1}A} \right){e^T}\left( t \right)  \\
\nonumber&+ 2{\Psi ^T}\left( t \right)\left( {{\Pi ^{\vartheta \left( t \right)}}{\Xi ^{\vartheta \left( t \right)}} \otimes P_2'^{ - 1}} \right)\tilde f\left( {x\left( t \right)} \right)  \\
\nonumber &- 2{e^T}\left( t \right)\left( {\left( {{\Pi ^{\vartheta \left( t \right)}}{\Xi ^{\vartheta \left( t \right)}}} \right)\left( {{L^{\vartheta \left( t \right)}}{\Gamma ^{\vartheta \left( t \right)}}} \right.} \right.\\
\nonumber  &+ \left. {{D^{\vartheta \left( t \right)}}{\Gamma ^{\vartheta \left( t \right)}}} \right) \otimes \left. {P_1'^{ - 1}BK} \right)\left( {e\left( t \right) + \Psi \left( t \right)} \right) \\
\nonumber&+ 2{e^T}\left( t \right)\left( {{\Pi ^{\vartheta \left( t \right)}}{\Xi ^{\vartheta \left( t \right)}} \otimes P_1'^{ - 1}} \right)\bar f\left( {x\left( t \right)} \right)   \\
\nonumber&+ 2{\Psi ^T}\left( t \right)\left( {{\Pi ^{\vartheta \left( t \right)}}{\Xi ^{\vartheta \left( t \right)}} \otimes \left( {P_2'^{ - 1}A + P_2'^{ - 1}GC} \right)} \right)\Psi \left( t \right) \\
\nonumber  \le & {{\mathord{\buildrel{\lower3pt\hbox{$\scriptscriptstyle\frown$}}
\over \varepsilon } }^T}\left( t \right)\left( {{I_N} \otimes \left( {A{P_1'} + {P_1'}{A^T} + {\rho ^2} + {P_1'}^T{P_1'}} \right)} \right)\mathord{\buildrel{\lower3pt\hbox{$\scriptscriptstyle\frown$}}
\over \varepsilon } \left( t \right)\\
\nonumber &- {{\mathord{\buildrel{\lower3pt\hbox{$\scriptscriptstyle\frown$}}
\over \varepsilon } }^T}\left( t \right)\left( {{I_N} \otimes \left( {\beta \left( {B{B^T} + \frac{1}{2}lB{B^T}} \right)} \right)} \right)\mathord{\buildrel{\lower3pt\hbox{$\scriptscriptstyle\frown$}}
\over \varepsilon } \left( t \right)\\
\nonumber &+ {{\mathord{\buildrel{\lower3pt\hbox{$\scriptscriptstyle\smile$}}
\over \Psi } }^T}\left( t \right)\left( {{I_N} \otimes \left( {A{P_2'} + {P_2'}{A^T} + 2GC{P_2'}} \right)} \right)\mathord{\buildrel{\lower3pt\hbox{$\scriptscriptstyle\smile$}}
\over \Psi } \left( t \right)\\
\nonumber &+ {{\mathord{\buildrel{\lower3pt\hbox{$\scriptscriptstyle\smile$}}
\over \Psi } }^T}\left( t \right)\left( {{I_N} \otimes \left( {{\rho ^2} + P_2'^T{P_2'}} \right)} \right)\mathord{\buildrel{\lower3pt\hbox{$\scriptscriptstyle\smile$}}
\over \Psi } \left( t \right)\\
 \nonumber &- {{\mathord{\buildrel{\lower3pt\hbox{$\scriptscriptstyle\smile$}}
\over \Psi } }^T}\left( t \right)\left( {{I_N} \otimes \left( {\frac{1}{2}\beta {l^{ - 1}}F{P_2'}{{\left( {F{P_2'}} \right)}^T}} \right)} \right)\mathord{\buildrel{\lower3pt\hbox{$\scriptscriptstyle\smile$}}
\over \Psi } \left( t \right)\\
\nonumber \le & - \beta {\lambda _{\min }}{\left( {P_1'^{ - 1},P_2'^{ - 1}} \right)^T}e\left( t \right)\left( {{\Pi ^{\vartheta \left( t \right)}}{\Xi ^{\vartheta \left( t \right)}} \otimes {I_N}} \right)e\left( t \right)\\
\nonumber  & - \beta {\lambda _{\min }}{\left( {P_2'^{ - 1},P_2'^{ - 1}} \right)^T}\Psi \left( t \right)\left( {{\Pi ^{\vartheta \left( t \right)}}{\Xi ^{\vartheta \left( t \right)}} \otimes {I_N}} \right)\Psi \left( t \right)\\
 \le & - \hat \gamma' V'\left( t \right)
\end{align}

When $\left[ {kw + \delta ,kw + h} \right) $, its time derivative as
\begin{align}
\nonumber \dot V'\left( t \right) =& 2{e^T}\left( t \right)\left( {{\Pi ^{\vartheta \left( t \right)}}{\Xi ^{\vartheta \left( t \right)}} \otimes P_1'^{ - 1}A} \right){e^T}\left( t \right)\\
\nonumber &- 2{e^T}\left( t \right)\left( {{\Pi ^{\vartheta \left( t \right)}}{\Xi ^{\vartheta \left( t \right)}}{L^{\vartheta \left( t \right)}}{\Gamma ^{\vartheta \left( t \right)}} \otimes P_1'^{ - 1}BK} \right)e\left( t \right)\\
\nonumber &- 2{e^T}\left( t \right)\left( {{\Pi ^{\vartheta \left( t \right)}}{\Xi ^{\vartheta \left( t \right)}}{L^{\vartheta \left( t \right)}}{\Gamma ^{\vartheta \left( t \right)}} \otimes P_1'^{ - 1}BK} \right)\Psi \left( t \right)\\
\nonumber & + 2{\Psi ^T}\left( t \right)\left( {{\Pi ^{\vartheta \left( t \right)}}{\Xi ^{\vartheta \left( t \right)}} \otimes \left( {P_2'^{ - 1}A + P_2'^{ - 1}GC} \right)} \right)\Psi \left( t \right)\\
\nonumber &+ 2{\Psi ^T}\left( t \right)\left( {{\Pi ^{\vartheta \left( t \right)}}{\Xi ^{\vartheta \left( t \right)}} \otimes P_2'^{ - 1}} \right)\tilde f\left( {x\left( t \right)} \right)\\
\nonumber  & + 2{e^T}\left( t \right)\left( {{\Pi ^{\vartheta \left( t \right)}}{\Xi ^{\vartheta \left( t \right)}} \otimes P_1'^{ - 1}} \right)\bar f\left( {x\left( t \right)} \right)\\
\nonumber  \le & {{\mathord{\buildrel{\lower3pt\hbox{$\scriptscriptstyle\frown$}}
\over \varepsilon } }^T}\left( t \right)\left( {{I_N} \otimes \left( {A{P_1'} + {P_1'}{A^T} + {\rho ^2} + {P_1'}^T{P_1'}} \right)} \right)\mathord{\buildrel{\lower3pt\hbox{$\scriptscriptstyle\frown$}}
\over \varepsilon } \left( t \right)\\
\nonumber &- {{\mathord{\buildrel{\lower3pt\hbox{$\scriptscriptstyle\frown$}}
\over \varepsilon } }^T}\left( t \right)\left( {{I_N} \otimes \left( {\beta \left( {B{B^T} + \frac{1}{2}lB{B^T}} \right)} \right)} \right)\mathord{\buildrel{\lower3pt\hbox{$\scriptscriptstyle\frown$}}
\over \varepsilon } \left( t \right)\\
\nonumber &+ {{\mathord{\buildrel{\lower3pt\hbox{$\scriptscriptstyle\smile$}}
\over \Psi } }^T}\left( t \right)\left( {{I_N} \otimes \left( {A{P_2'} + {P_2'}{A^T} + 2GC{P_2'}} \right)} \right)\mathord{\buildrel{\lower3pt\hbox{$\scriptscriptstyle\smile$}}
\over \Psi } \left( t \right)\\
 \nonumber &- {{\mathord{\buildrel{\lower3pt\hbox{$\scriptscriptstyle\smile$}}
\over \Psi } }^T}\left( t \right)\left( {{I_N} \otimes \left( {\frac{1}{2}\beta {l^{ - 1}}F{P_2'}{{\left( {F{P_2'}} \right)}^T}} \right)} \right)\mathord{\buildrel{\lower3pt\hbox{$\scriptscriptstyle\smile$}}
\over \Psi } \left( t \right)\\
\nonumber &+ {{\mathord{\buildrel{\lower3pt\hbox{$\scriptscriptstyle\smile$}}
\over \Psi } }^T}\left( t \right)\left( {{I_N} \otimes \left( {{\rho ^2} + P_2'^T{P_2'}} \right)} \right)\mathord{\buildrel{\lower3pt\hbox{$\scriptscriptstyle\smile$}}
\over \Psi } \left( t \right)\\
\nonumber  \le & {\lambda _{\max }}{\left( {{M^3},{M^4}} \right)^T}\left( {{{\mathord{\buildrel{\lower3pt\hbox{$\scriptscriptstyle\frown$}}
\over \varepsilon } }^T}\left( t \right)\mathord{\buildrel{\lower3pt\hbox{$\scriptscriptstyle\frown$}}
\over \varepsilon } \left( t \right) + {{\mathord{\buildrel{\lower3pt\hbox{$\scriptscriptstyle\smile$}}
\over \Psi } }^T}\left( t \right)\mathord{\buildrel{\lower3pt\hbox{$\scriptscriptstyle\smile$}}
\over \Psi } \left( t \right)} \right)\\
 \le & \hat l'V'\left( t \right)
\end{align}

For ${t \in \left[ {kw + h ,\left( {k + 1} \right)w} \right)}$, we have
\begin{align}
\nonumber
\dot V'\left( t \right) =& 2{e^T}\left( t \right)\left( {{\Pi ^{\vartheta \left( t \right)}}{\Xi ^{\vartheta \left( t \right)}} \otimes P_1'^{ - 1}A} \right){e^T}\left( t \right)\\
\nonumber &+ 2{\Psi ^T}\left( t \right)\left( {{\Pi ^{\vartheta \left( t \right)}}{\Xi ^{\vartheta \left( t \right)}} \otimes P_2'^{ - 1}} \right)\tilde f\left( {x\left( t \right)} \right)\\
\nonumber &+ 2{e^T}\left( t \right)\left( {{\Pi ^{\vartheta \left( t \right)}}{\Xi ^{\vartheta \left( t \right)}} \otimes P_1'^{ - 1}} \right)\bar f\left( {x\left( t \right)} \right)\\
\nonumber & + 2{\Psi ^T}\left( t \right)\left( {{\Pi ^{\vartheta \left( t \right)}}{\Xi ^{\vartheta \left( t \right)}} \otimes \left( {P_2'^{ - 1}A + P_2'^{ - 1}GC} \right)} \right)\Psi \left( t \right) \\
\nonumber \le & {{\mathord{\buildrel{\lower3pt\hbox{$\scriptscriptstyle\frown$}}
\over \varepsilon } }^T}\left( t \right)\left( {{I_N} \otimes \left( {A{P_1'} + {P_1'}{A^T}+{\rho ^2}} \right)} \right)\mathord{\buildrel{\lower3pt\hbox{$\scriptscriptstyle\frown$}}
\over \varepsilon } \left( t \right)\\
\nonumber  & + {{\mathord{\buildrel{\lower3pt\hbox{$\scriptscriptstyle\smile$}}
\over \Psi } }^T}\left( t \right)\left( {{I_N} \otimes \left( {A{P_2'} + {P_2'}{A^T} + {\rho ^2}} \right)} \right)\mathord{\buildrel{\lower3pt\hbox{$\scriptscriptstyle\smile$}}
\over \Psi } \left( t \right)\\
\nonumber   & + {{\mathord{\buildrel{\lower3pt\hbox{$\scriptscriptstyle\smile$}}
\over \Psi } }^T}\left( t \right)\left( {{I_N} \otimes \left( {2GC{P_2'} + P_2'^T{P_2'}} \right)} \right)\mathord{\buildrel{\lower3pt\hbox{$\scriptscriptstyle\smile$}}
\over \Psi } \left( t \right)\\
\nonumber &+ {{\mathord{\buildrel{\lower3pt\hbox{$\scriptscriptstyle\frown$}}
\over \varepsilon } }^T}\left( t \right)\left( {{I_N} \otimes \left( {{P_1'}^T{P_1'} } \right)} \right)\mathord{\buildrel{\lower3pt\hbox{$\scriptscriptstyle\frown$}}
\over \varepsilon } \left( t \right)\\
\nonumber \le & {\lambda _{\max }}{\left( {{M^5},{M^6}} \right)^T}\left( {{{\mathord{\buildrel{\lower3pt\hbox{$\scriptscriptstyle\frown$}}
\over \varepsilon } }^T}\left( t \right)\mathord{\buildrel{\lower3pt\hbox{$\scriptscriptstyle\frown$}}
\over \varepsilon } \left( t \right) + {{\mathord{\buildrel{\lower3pt\hbox{$\scriptscriptstyle\smile$}}
\over \Psi } }^T}\left( t \right)\mathord{\buildrel{\lower3pt\hbox{$\scriptscriptstyle\smile$}}
\over \Psi } \left( t \right)} \right)\\
\le & {\mathord{\buildrel{\lower3pt\hbox{$\scriptscriptstyle\frown$}}
\over \chi } } V'\left( t \right)
\end{align}

where ${\mathord{\buildrel{\lower3pt\hbox{$\scriptscriptstyle\frown$}}
\over \chi } } = {\lambda _{\max }}\left( {{M^5},{M^6}} \right){\lambda _{\max }}\left( {P_1'^{ - 1},P_2'^{ - 1}} \right)$.
Note that the cooperative
constrained weighting $\hat l' $ is vary in $[{ - \beta ,\mathord{\buildrel{\lower3pt\hbox{$\scriptscriptstyle\frown$}}
\over \chi } }]$ and that depends
on the solution of (39) and (40). Thus, we can get
\begin{align}
\nonumber V'\left( w \right) \le & V'\left( h \right){e^{\hat l'\left( {w - h} \right)}}\\
 \le & V'\left( \delta  \right){e^{\mathord{\buildrel{\lower3pt\hbox{$\scriptscriptstyle\frown$}}
\over \chi } \left( {h - \delta } \right)}}{e^{\hat l'\left( {w - h} \right)}}\\
\nonumber  \le & V'\left( 0 \right){e^{ - \hat \gamma' \delta  + \hat l'\left( {w - h - \delta } \right) + \mathord{\buildrel{\lower3pt\hbox{$\scriptscriptstyle\frown$}}
\over \chi } \left( {h - \delta } \right)}}\\
 = & V'\left( 0 \right){e^{ - {{\mathord{\buildrel{\lower3pt\hbox{$\scriptscriptstyle\frown$}}
\over \Upsilon } }_1}}}
\end{align}
where ${{\mathord{\buildrel{\lower3pt\hbox{$\scriptscriptstyle\frown$}}
\over \Upsilon } }_1} = \hat \gamma' \delta  - \hat l'\left( {w - h - \delta } \right) - \mathord{\buildrel{\lower3pt\hbox{$\scriptscriptstyle\frown$}}
\over \chi } \left( {h - \delta } \right)$. The range of ${{\mathord{\buildrel{\lower3pt\hbox{$\scriptscriptstyle\frown$}}
\over \Upsilon } }_1}$ can be obtained by (45), that is ${{\mathord{\buildrel{\lower3pt\hbox{$\scriptscriptstyle\frown$}}
\over \Upsilon } }_1}> 0$. For any positive integer $k$, the following equation holds
\begin{align}
V'\left( {\left( {k + 1} \right)w} \right) \le V'\left( 0 \right){e^{ - \sum\limits_{j = 1}^k {{{{{\mathord{\buildrel{\lower3pt\hbox{$\scriptscriptstyle\frown$}}
\over \Upsilon } }}}_j}} }}
\end{align}
where ${{\mathord{\buildrel{\lower3pt\hbox{$\scriptscriptstyle\frown$}}
\over \Upsilon } }_j} = \hat \gamma' {\delta _j} - \hat l'\left( {w - {h_j} - {\delta _j}} \right) - \mathord{\buildrel{\lower3pt\hbox{$\scriptscriptstyle\frown$}}
\over \chi } \left( {{h_j} - {\delta _j}} \right)$, $j = 1, \ldots ,k$. Set $2sw \le t \le \left( {2s + 1} \right)w$ for any $s \in R $. Define $\left[ {kw,(k + 1)w} \right)$, $k \in {{\rm{N}}_ + }$  is a uniformly bounded time sequence, ${\mathord{\buildrel{\lower3pt\hbox{$\scriptscriptstyle\smile$}}
\over \omega } _{\max }} = {\max _{k \in {_ + }}}\left\{ {\left( {k + 1} \right)w,kw} \right\}$ and ${\mathord{\buildrel{\lower3pt\hbox{$\scriptscriptstyle\frown$}}
\over \Upsilon } }  = {\min _{j \in {{\rm{N}}_ + }}}\left\{ {{{{\mathord{\buildrel{\lower3pt\hbox{$\scriptscriptstyle\frown$}}
\over \Upsilon } } }_j}} \right\}$. Finally, according to the recursion principle, we can obtain
\begin{align}
\nonumber
V'\left( t \right) & \le V'\left( {2sw} \right){e^{{{\mathord{\buildrel{\lower3pt\hbox{$\scriptscriptstyle\smile$}}
\over \omega } }_{\max }}(\hat \gamma'  + \hat l' + \mathord{\buildrel{\lower3pt\hbox{$\scriptscriptstyle\frown$}}
\over \chi } )}}\\
\nonumber & \le {e^{{{\mathord{\buildrel{\lower3pt\hbox{$\scriptscriptstyle\smile$}}
\over \omega } }_{\max }}(\hat \gamma'  + \hat l' + \mathord{\buildrel{\lower3pt\hbox{$\scriptscriptstyle\frown$}}
\over \chi } )}}V'\left( 0 \right){e^{ - \sum\nolimits_{j = 1}^s {{{{\mathord{\buildrel{\lower3pt\hbox{$\scriptscriptstyle\frown$}}
\over \Upsilon } } }_j}} }}\\
& \le  {\Omega _0'}{e^{ - {\Omega _1'}t}}
\end{align}
where ${\Omega _0'} = {e^{{{\mathord{\buildrel{\lower3pt\hbox{$\scriptscriptstyle\smile$}}
\over \omega } }_{\max }}(\hat \gamma'  + \hat l' + \mathord{\buildrel{\lower3pt\hbox{$\scriptscriptstyle\frown$}}
\over \chi } )}}{V}'\left( 0 \right)$ and ${\Omega _1} = {{{\mathord{\buildrel{\lower3pt\hbox{$\scriptscriptstyle\frown$}}
\over \Upsilon } } } \mathord{\left/
 {\vphantom {{\tilde \Upsilon } {{{\mathord{\buildrel{\lower3pt\hbox{$\scriptscriptstyle\smile$}}
\over \omega } }_{\max }}}}} \right.
 \kern-\nulldelimiterspace} {{{\mathord{\buildrel{\lower3pt\hbox{$\scriptscriptstyle\smile$}}
\over \omega } }_{\max }}}}$.

This completes the proof. \end{IEEEproof}

	\section{Numerical Examples}

In this section, two simulation examples are provided to verify the effectiveness of the obtained results.
Both nonlinear MASs
consist of four followers and a leader, with the directed  dynamic interaction topology as shown in Fig. 1, where 0 is
the leader and 1-4 are the followers.
\begin{figure}[thpb]
		\centering
		\includegraphics[width=9cm]{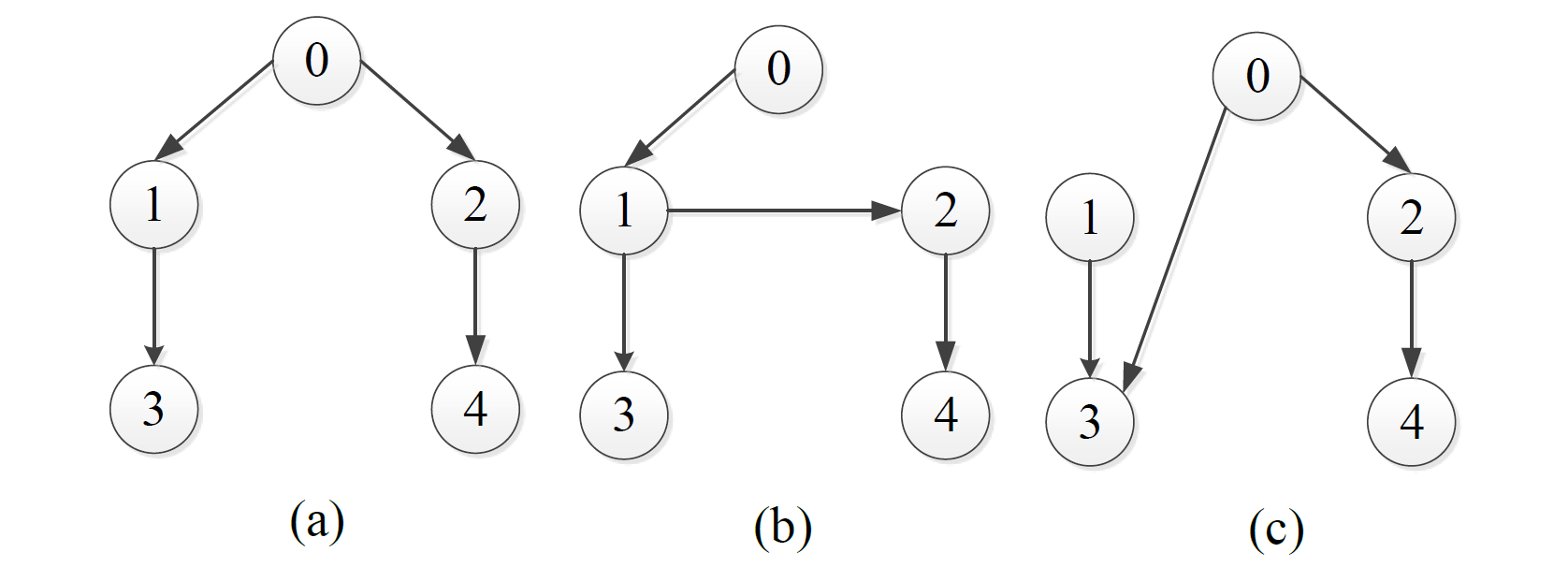}
		\caption{{Three dynamic interaction topologies between the leader and followers: (a) ${\tilde G_1}$, (b) ${\tilde G_2}$, (c) ${\tilde G_3}$ } }
		\label{}
	\end{figure}

The dynamic of the $ith$ agent for nonlinear MASs (1) can be written with
\begin{equation*}
{x_i}\left( t \right) = \left[ {\begin{array}{*{20}{c}}
{{x_{i1}}\left( t \right)}\\
{{x_{i2}}\left( t \right)}\\
{{x_{i3}}\left( t \right)}\\
{{x_{i4}}\left( t \right)}
\end{array}} \right],A = \left[ {\begin{array}{*{20}{c}}
0&1&0&0\\
{ - 48.6}&{ - 1.25}&{48.6}&0\\
0&0&0&1\\
{1.95}&0&{ - 1.95}&0
\end{array}} \right]
\end{equation*}

\begin{equation*}
B = \left[ {\begin{array}{*{20}{c}}
0\\
{21.6}\\
0\\
0
\end{array}} \right],C = \left[ {\begin{array}{*{20}{c}}
1&0&0&0\\
0&0&1&0
\end{array}} \right]
\end{equation*}

The nonlinear dynamic functions are:
\begin{align*}
f\left( {{x_i}\left( t \right)} \right) &= {\left[ {\begin{array}{*{20}{c}}
0&0&0&{ - 3.33\sin ({x_{i3}})}
\end{array}} \right]^T}\\
f\left( {{x_0}\left( t \right)} \right) &= {\left[ {\begin{array}{*{20}{c}}
0&0&0&{ - 3.33\sin ({x_{03}})}
\end{array}} \right]^T},\forall i = 1,2, \ldots ,n
\end{align*}

\begin{figure}[thpb]
		\centering
		\includegraphics[width=8cm]{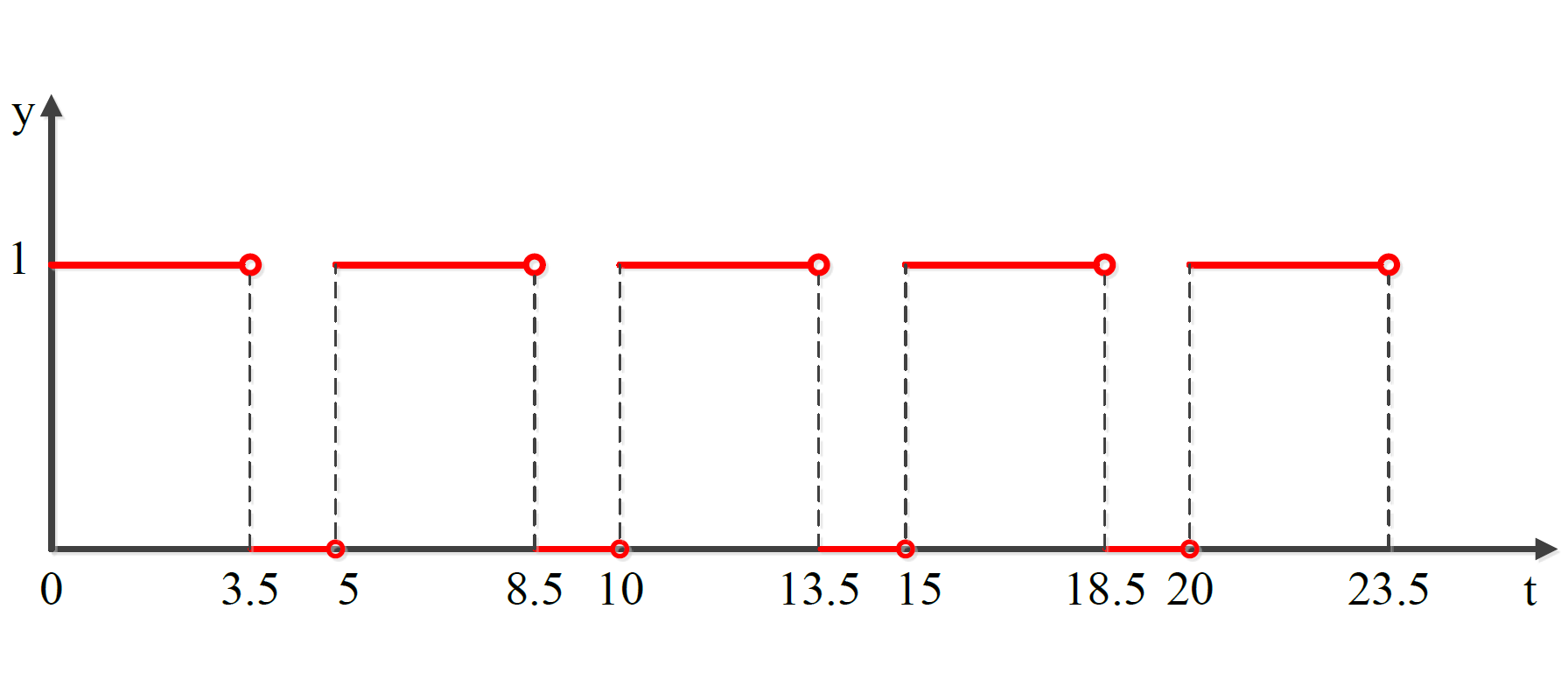}
		\caption{{  Intermittent communication } }
		\label{}
	\end{figure}

Consider the nonlinear MASs (1) compose of four followers and a leader, whose dynamic switching communication network is shown in Fig. 1, where node 0 represents the leader and the remaining nodes 1-4 are the followers.
For the convenience of analysis, suppose that the underlying derivative directed network topologies can be constructed as shown in Fig. 1(a), (b) and (c).

\begin{figure}[thpb]
		\centering
		\includegraphics[width=9cm]{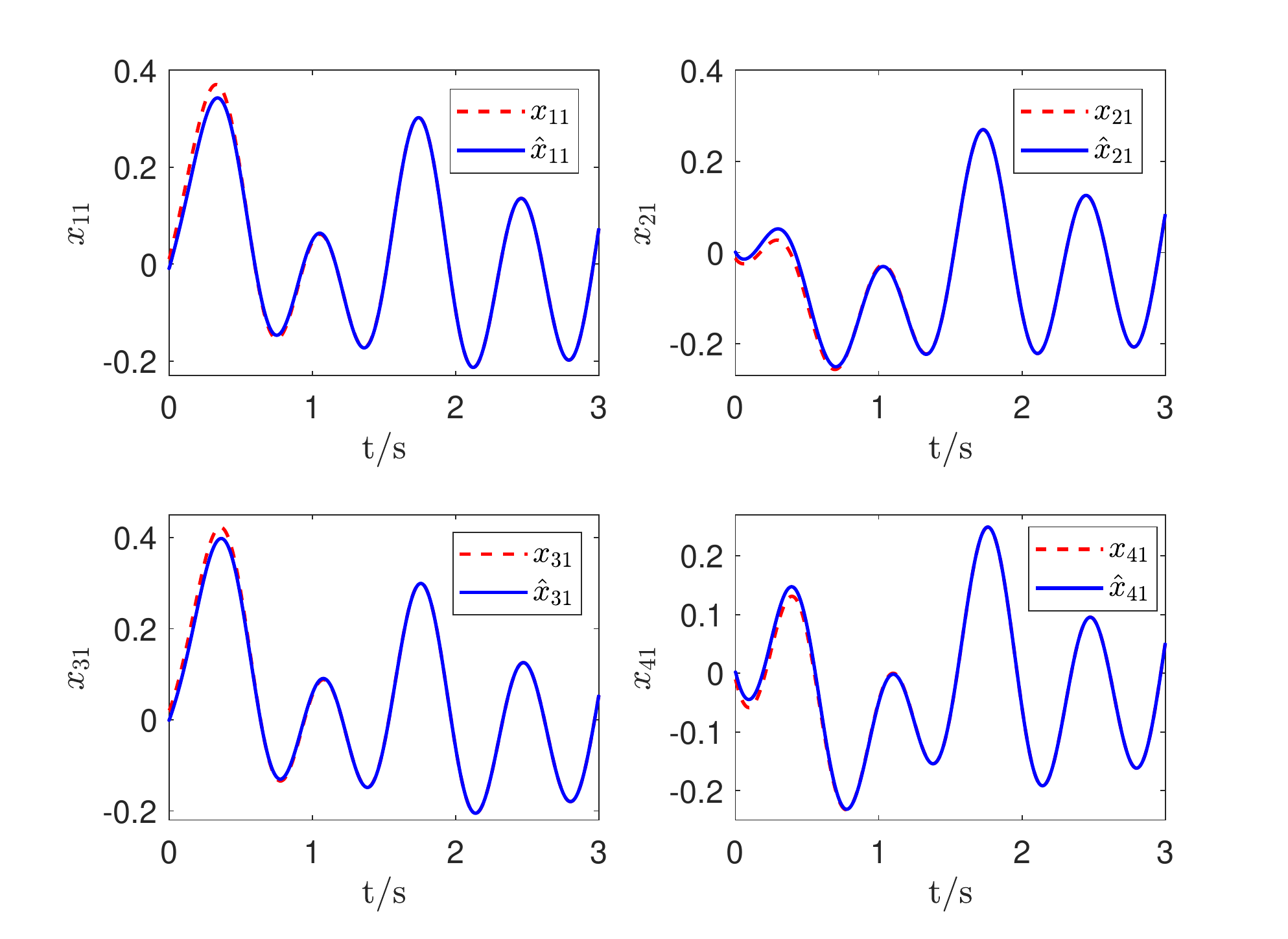}
		\caption{Actual and estimated values under the designed observer (3) for the first state}
		\label{}
\end{figure}

\begin{figure}[thpb]
		\centering
		\includegraphics[width=9cm]{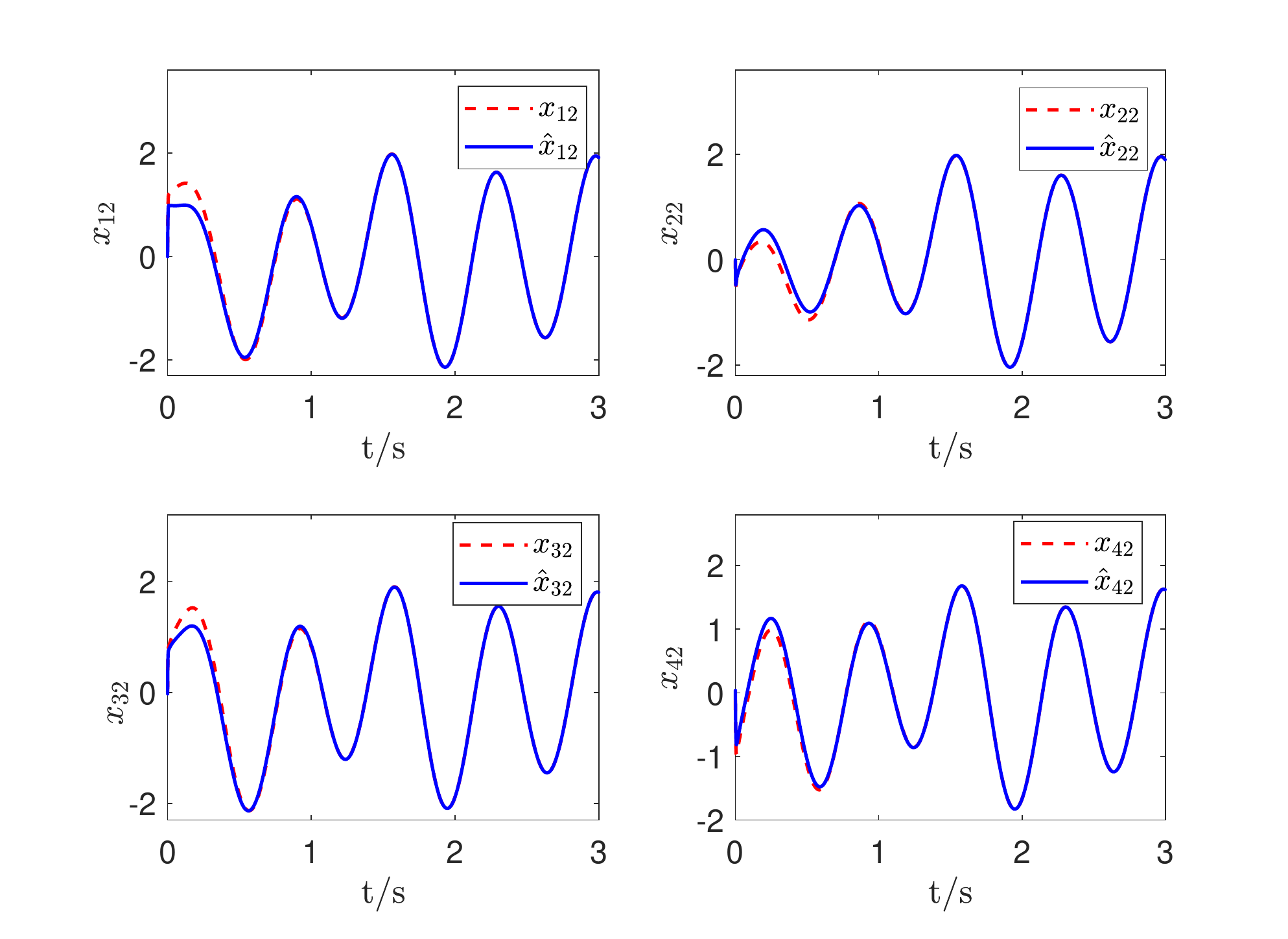}
		\caption{Actual and estimated values under the designed observer (3) for the second state}
		\label{}
	\end{figure}
\begin{figure}[thpb]
		\centering
		\includegraphics[width=9cm]{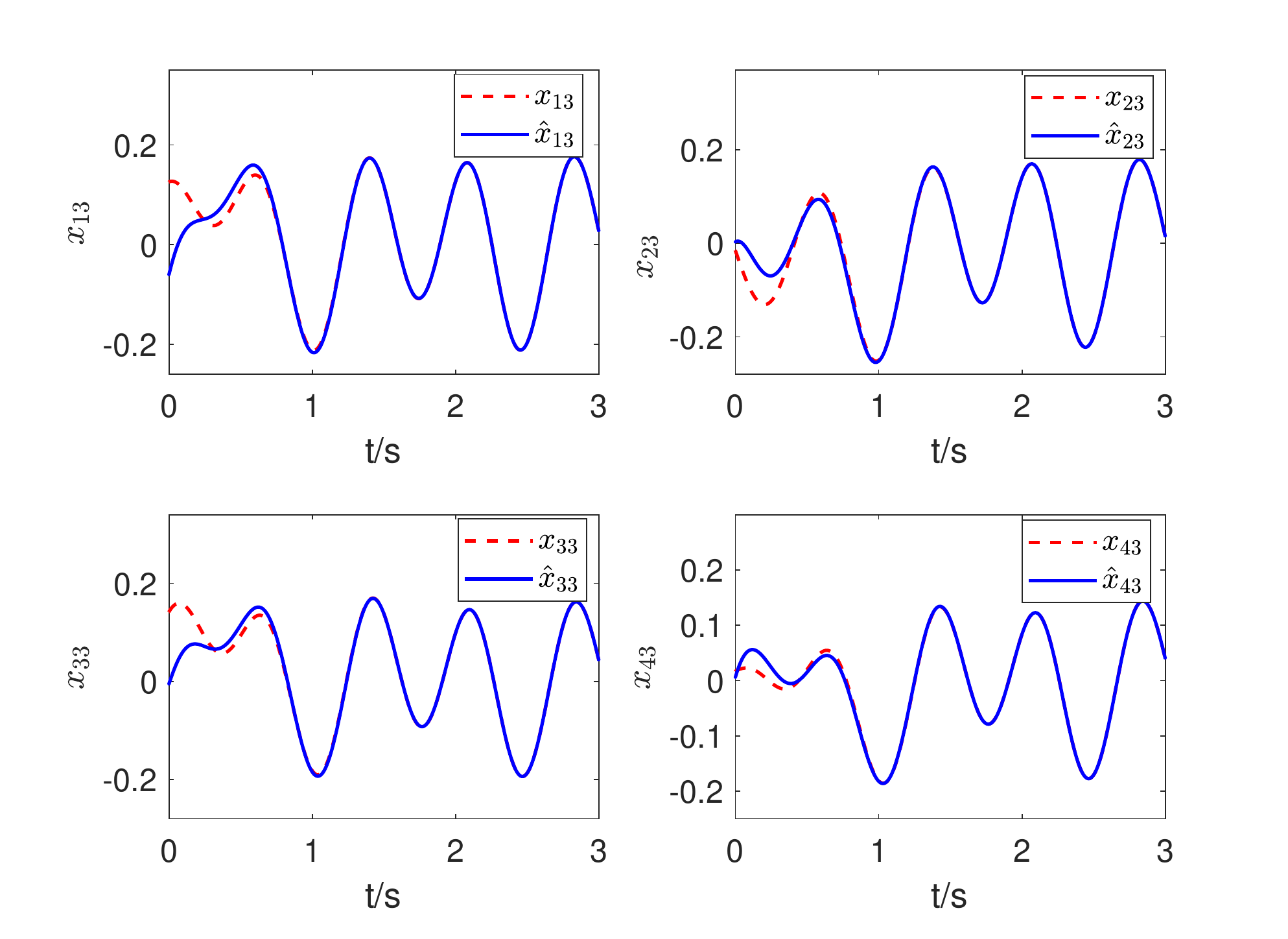}
		\caption{Actual and estimated values under the designed observer (3) for the third state}
		\label{}
	\end{figure}

\begin{figure}[thpb]
		\centering
		\includegraphics[width=9cm]{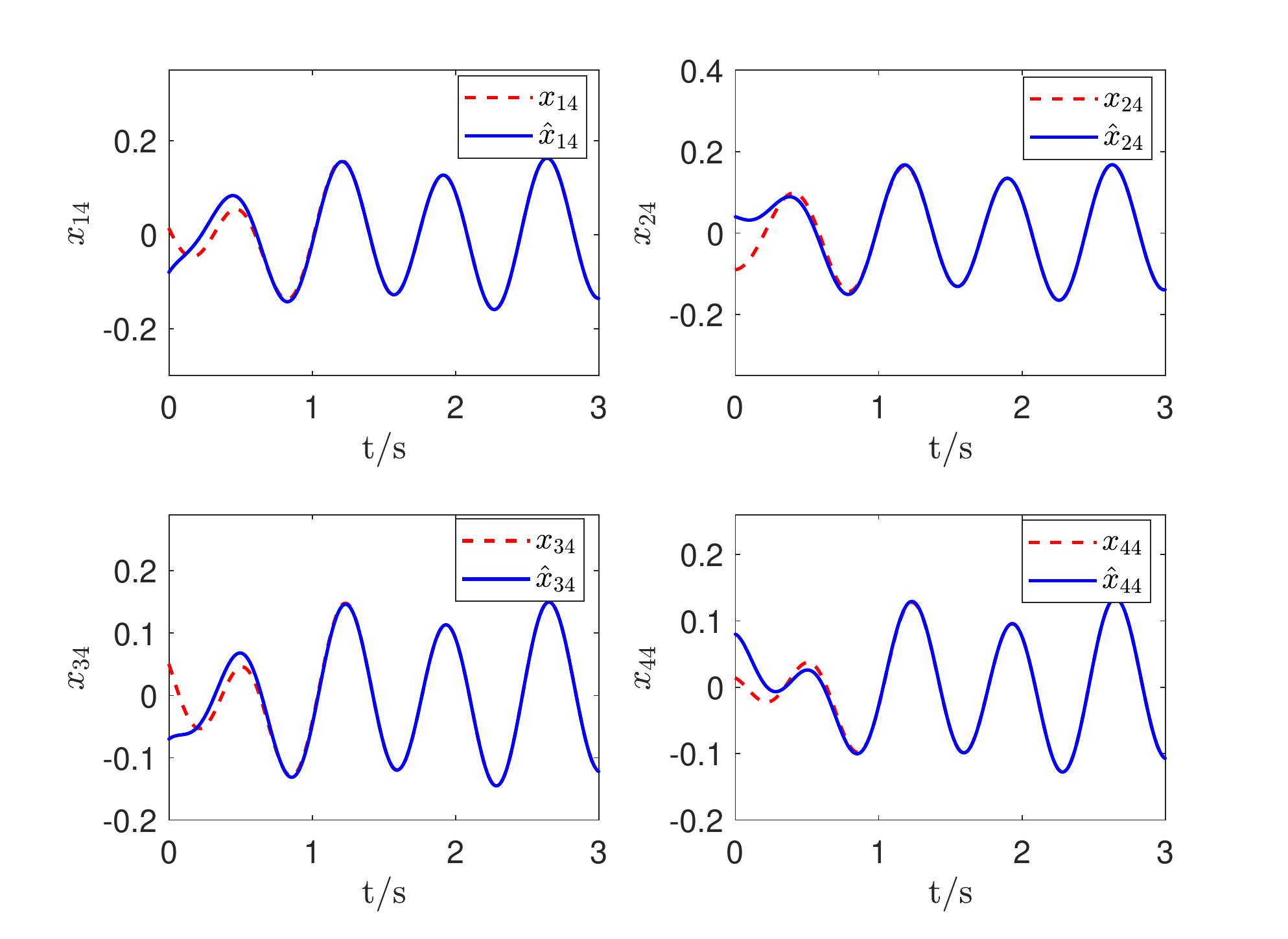}
		\caption{Actual and estimated values under the designed observer (3) for the fourth state}
		\label{}
	\end{figure}

\emph{Example 1}: First, we consider the intermittent constrained consensus for cooperative tracking control. The intermittent communication mode is shown in Fig.2, in which, if y=1, it indicates that the communication between adjacent agents in the system is working, and y=0 shows the neighboring agents can not communicate with each other.

According to Theorem 1, there exists an infinite time sequence of uniformly bounded and nonoverlapping time intervals $t \in \left[ {kw,(k + 1)w} \right)$, where $w = 5$. To further analyze the stability, the switching sequence within the time interval $t \in \left[ {kw,kw + \delta } \right)$ is defined as ${\tilde G_1} \to {\tilde G_2} \to {\tilde G_3} \to {\tilde G_1} \to {\tilde G_2} \to {\tilde G_3}$. The switching signal is shown as Fig. 1.
Assuming the communication duration $\delta  = 3.5$ and the control parameters $\beta  = 0.01, l = 0.02, \rho  = 0.2$.
According to Lemma 5 and distributed coordinated tracking controller (3), the positive definite matrices ${P_1}$, ${P_2}$, the controller feedback gain matrix and observer gain matrix can be obtained as:
\begin{align*}
{P_1} &= \left[ {\begin{array}{*{20}{c}}
{0.39}&{ - 0.36}&{0.36}&{ - 0.04}\\
{ - 0.36}&{2.03}&{ - 0.32}&{ - 0.01}\\
{0.36}&{ - 0.32}&{0.38}&{ - 0.3}\\
{ - 0.04}&{ - 0.01}&{ - 0.3}&{0.01}
\end{array}} \right]\\
{P_2} &= \left[ {\begin{array}{*{20}{c}}
{7.33}&{ - 0.09}&{0.02}&{ - 0.96}\\
{ - 0.09}&{7.47}&{0.01}&{ - 0.46}\\
{0.02}&{0.01}&{0.38}&{0.08}\\
{ - 0.96}&{ - 0.46}&{0.08}&{2.49}
\end{array}} \right]\\
K &= \left[ {\begin{array}{*{20}{c}}
{ - 18.3}&{ - 8.75}&{1.58}&{ - 95.48}
\end{array}} \right] \\
\bar G &= \left[ {\begin{array}{*{20}{c}}
{ - 10.97}&{0.67}\\
{23.38}&{ - 23.96}\\
{ - 0.46}&{ - 10.66}\\
{ - 5.38}&{ - 0.30}
\end{array}} \right]
\end{align*}

The trajectories of the actual states and estimates states of four followers are provided in Figs. 3-6, respectively for first state, second state, third state and fourth state.
It can be proved that the constructed observers (3) can estimate the actual states asymptotically.
The state trajectories of the closed-loop MASs (1) under the heterogeneous coupling framework protocol (4) are shown in Figs. 7-10 show that the coordinated tracking control with intermittent communication and dynamic switching topology is indeed achieved.

\begin{figure}[thpb]
		\centering
		\includegraphics[width=9cm]{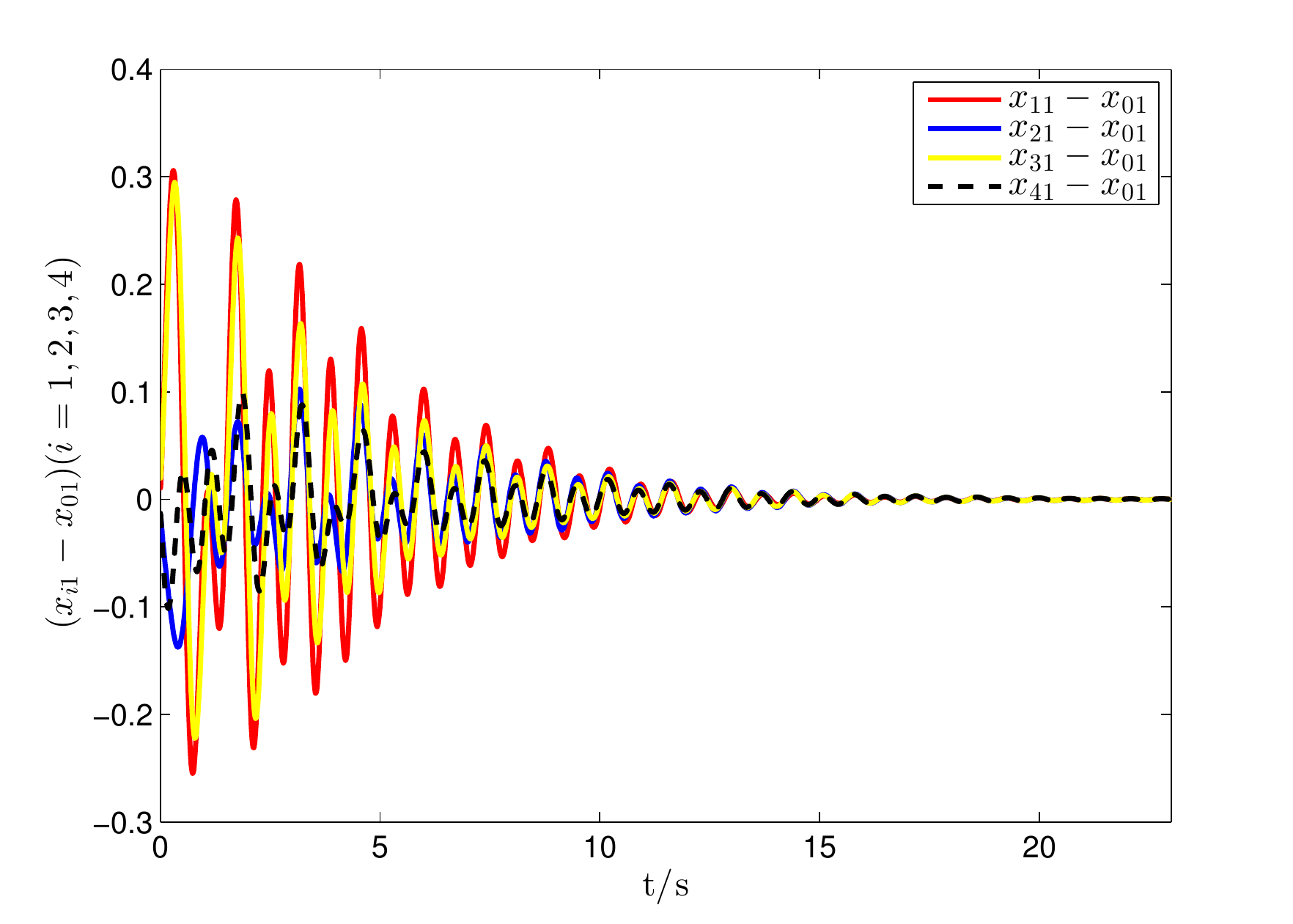}
		\caption{The first state errors for four followers and the leader under the designed controller (4).}
		\label{}
		\centering
		\includegraphics[width=9cm]{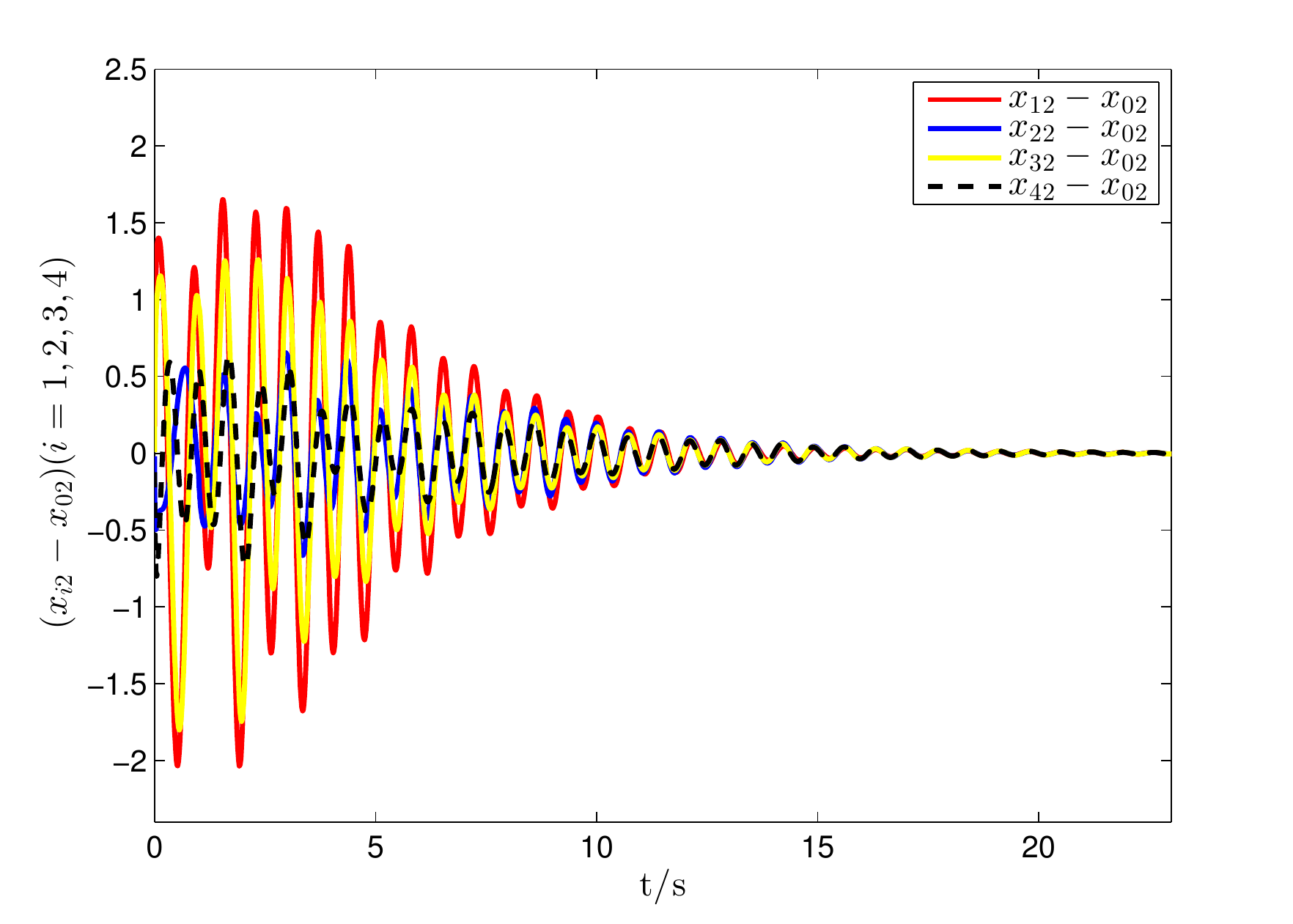}
		\caption{The second state errors for four followers and the leader under the designed controller (4).}
		\label{}
		\centering
		\includegraphics[width=9cm]{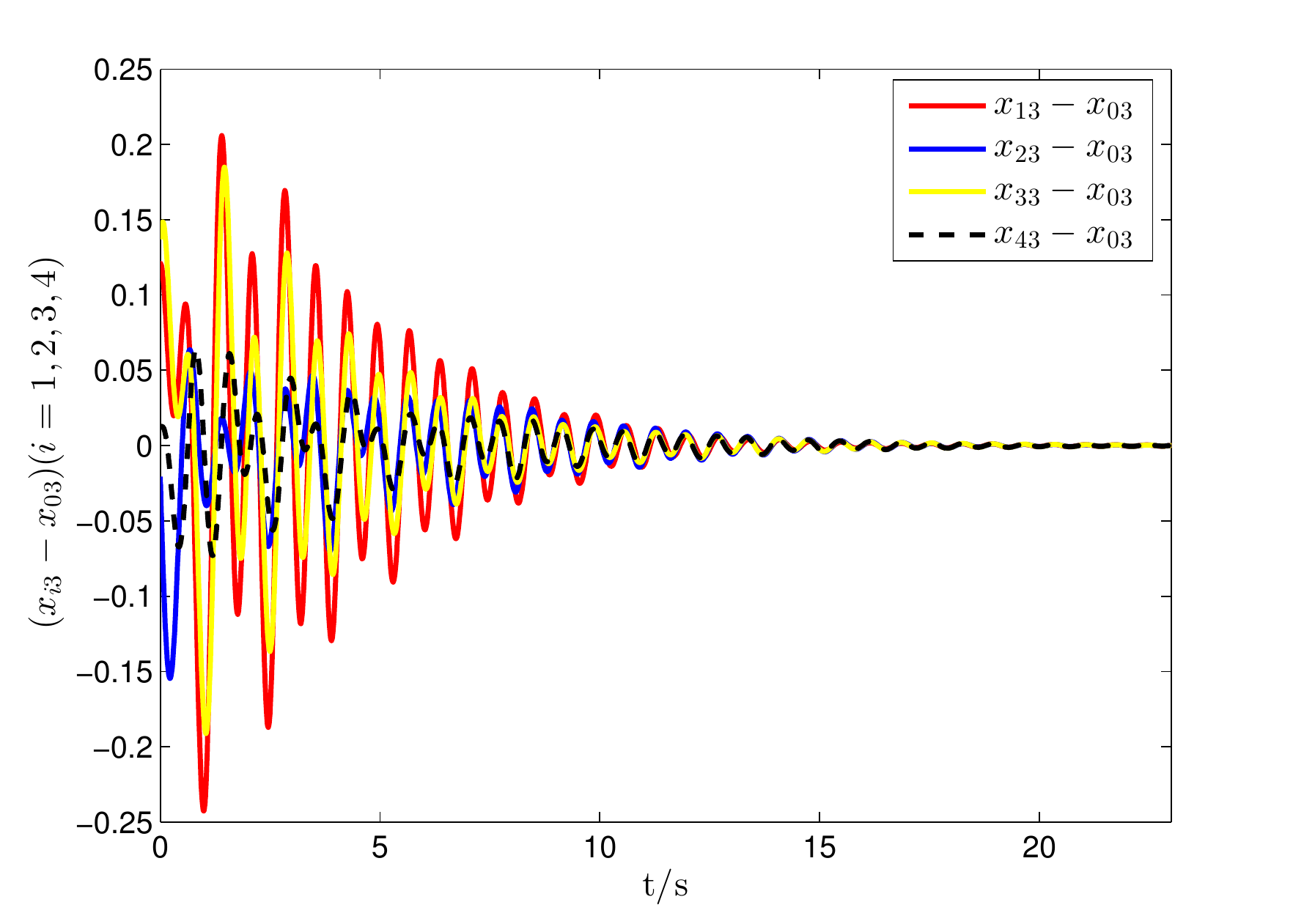}
		\caption{The third state errors for four followers and the leader under the designed controller (4).}
		\label{}
	\end{figure}

\begin{figure}[thpb]
		\centering
		\includegraphics[width=9cm]{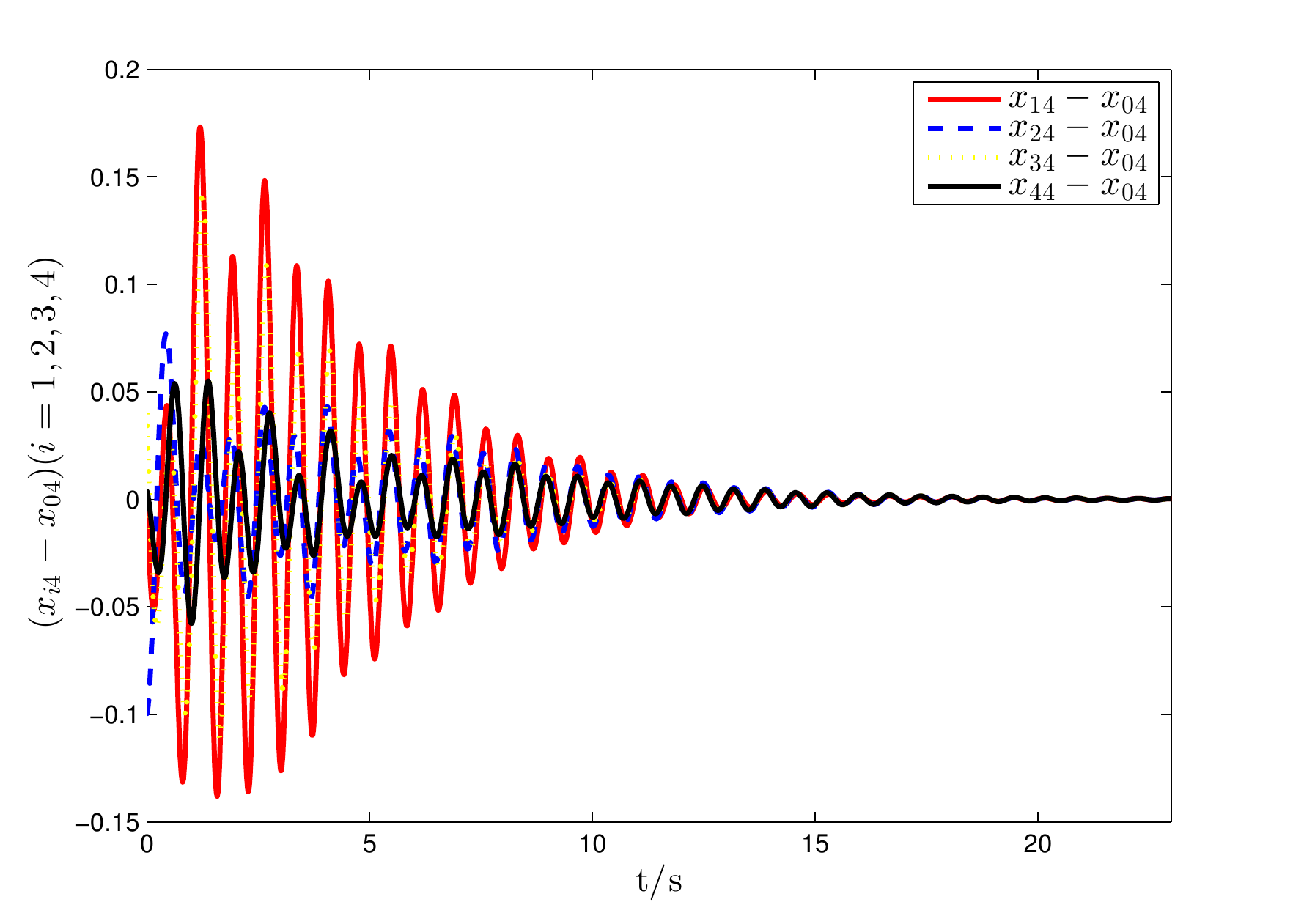}
		\caption{The fourth state errors for four followers and the leader under the designed controller (4).}
		\label{}
	\end{figure}

	\emph{Example 2}:	Second, we consider the observer-based consensus control under second intermittent coordinated constrained with three types of intervals.
The intermittent communication mode is shown in Fig. 11, in which, if y=2, it indicates that the communication between adjacent agents and the leader in the system is working, if y=1, it indicates that each follower evolves only based on the information of itself and its neighbors,
and y=0 shows the neighboring agents can not communicate with each other.

Three types of time intervals $t = \left[ {5k,5k+ 4} \right)=T^m$, $t = \left[ {5k,5k+ 4.5} \right)=T^q$  and $t = \left[ {5k + 4.5,5k + 5} \right)=T^n$
respectively are considered for simulation.

	\begin{figure}[thpb]
		\centering
		\includegraphics[width=9cm]{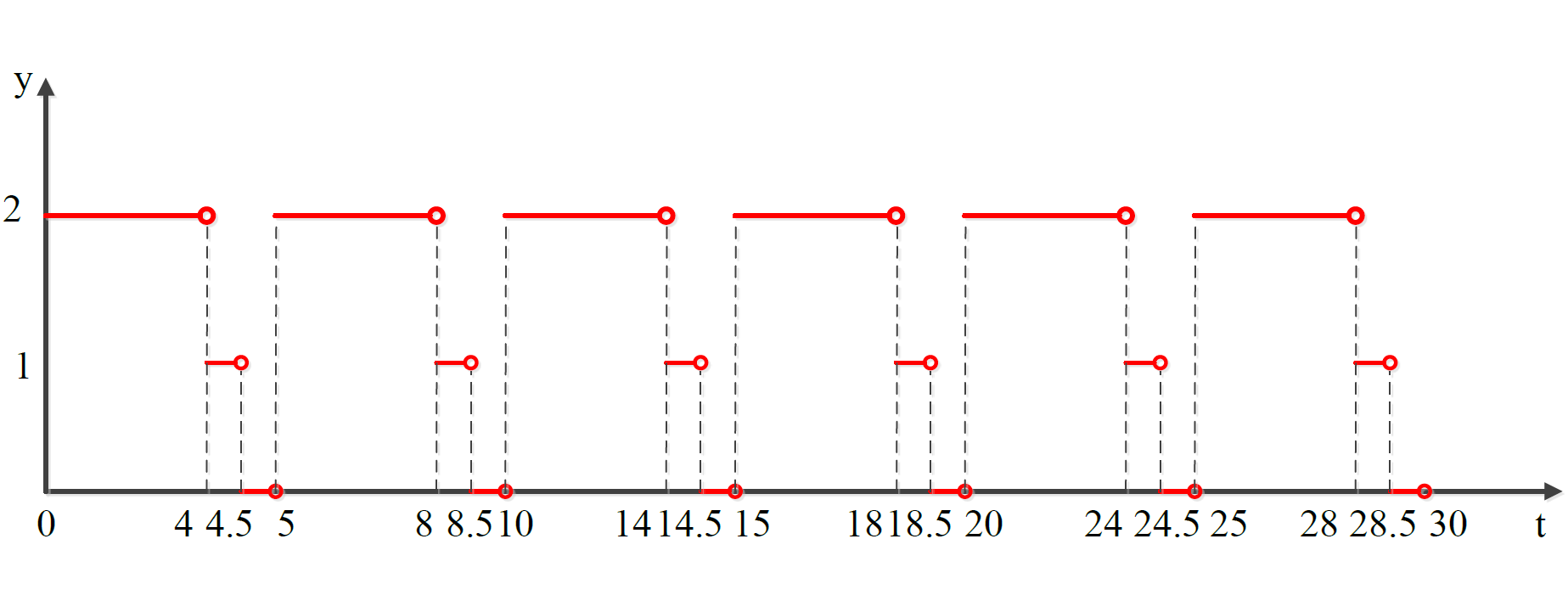}
		\caption{Intermittent Coordinated Constrained}
		\label{}
	\end{figure}

\begin{figure}[thpb]
		\centering
		\includegraphics[width=9cm]{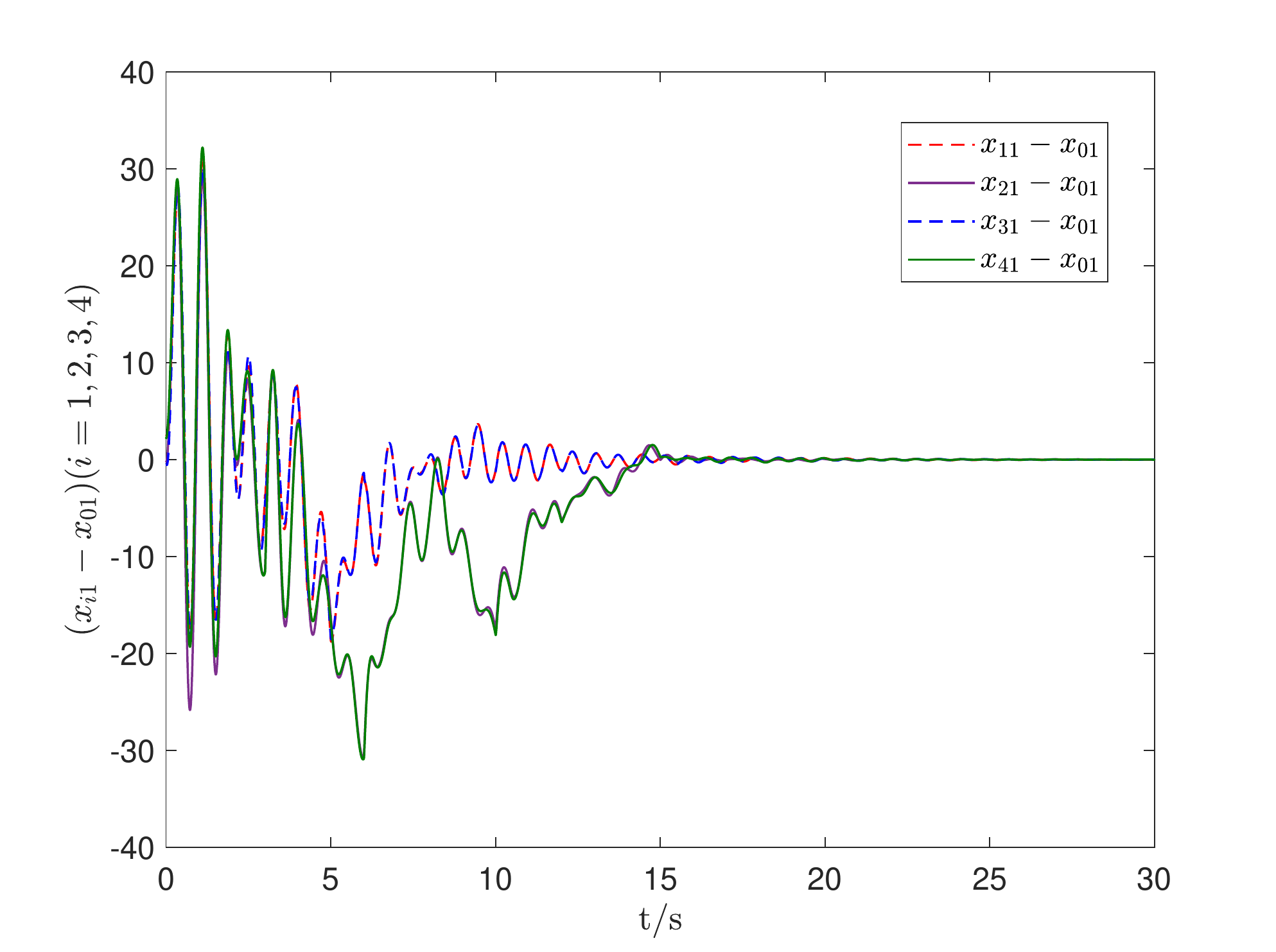}
		\caption{Tracking errors with $x_{i1}$ and $x_{01}$  under the designed controller (38)}
		\label{}
	\end{figure}

	\begin{figure}[thpb]
		\centering
		\includegraphics[width=9cm]{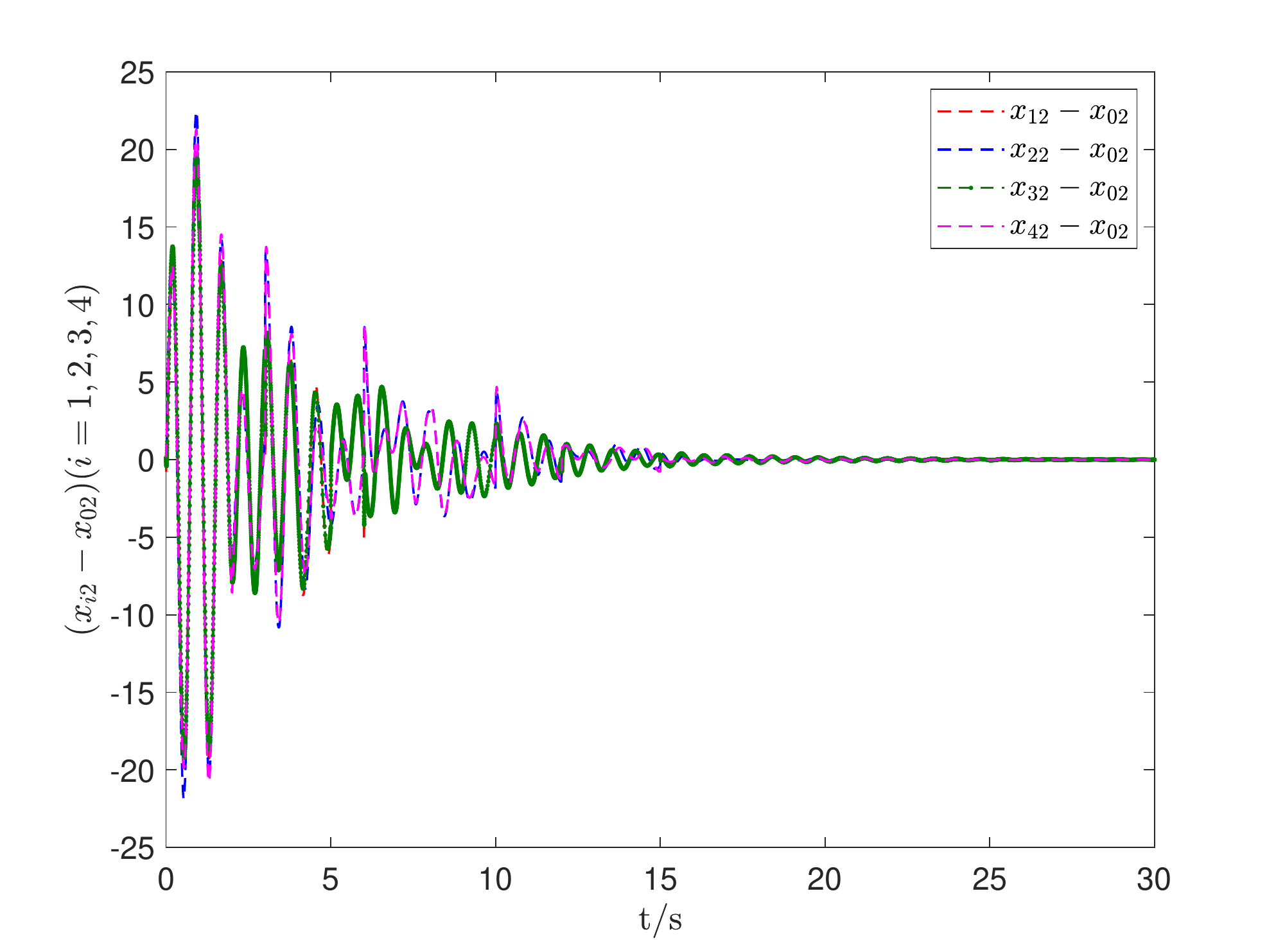}
		\caption{Tracking errors with $x_{i2}$ and $x_{02}$  under the designed controller (38)}
		\label{}
		\centering
		\includegraphics[width=9cm]{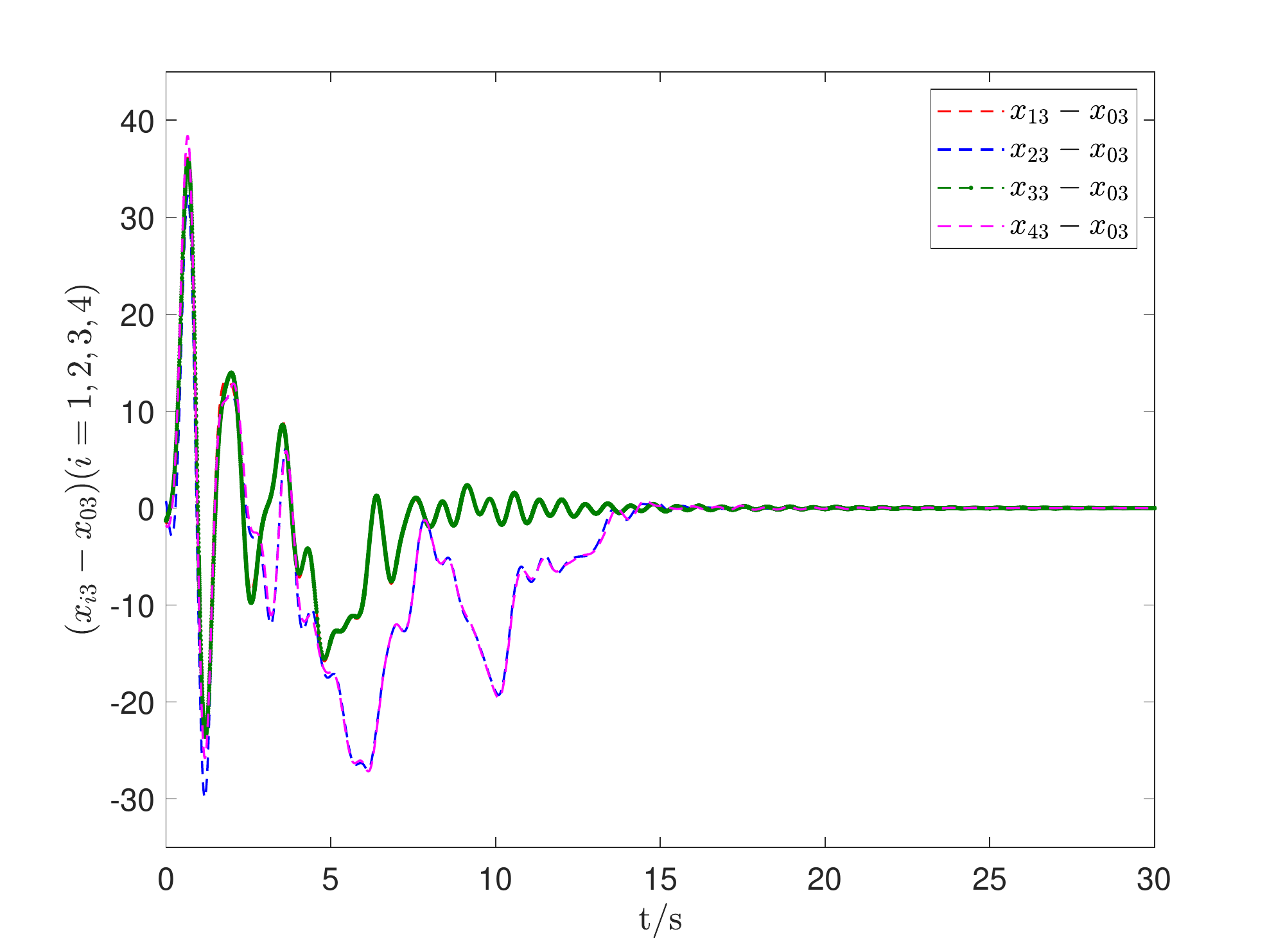}
		\caption{Tracking errors with $x_{i3}$ and $x_{03}$  under the designed controller (38)}
		\label{}
	\end{figure}

	\begin{figure}[thpb]
		\centering
		\includegraphics[width=9cm]{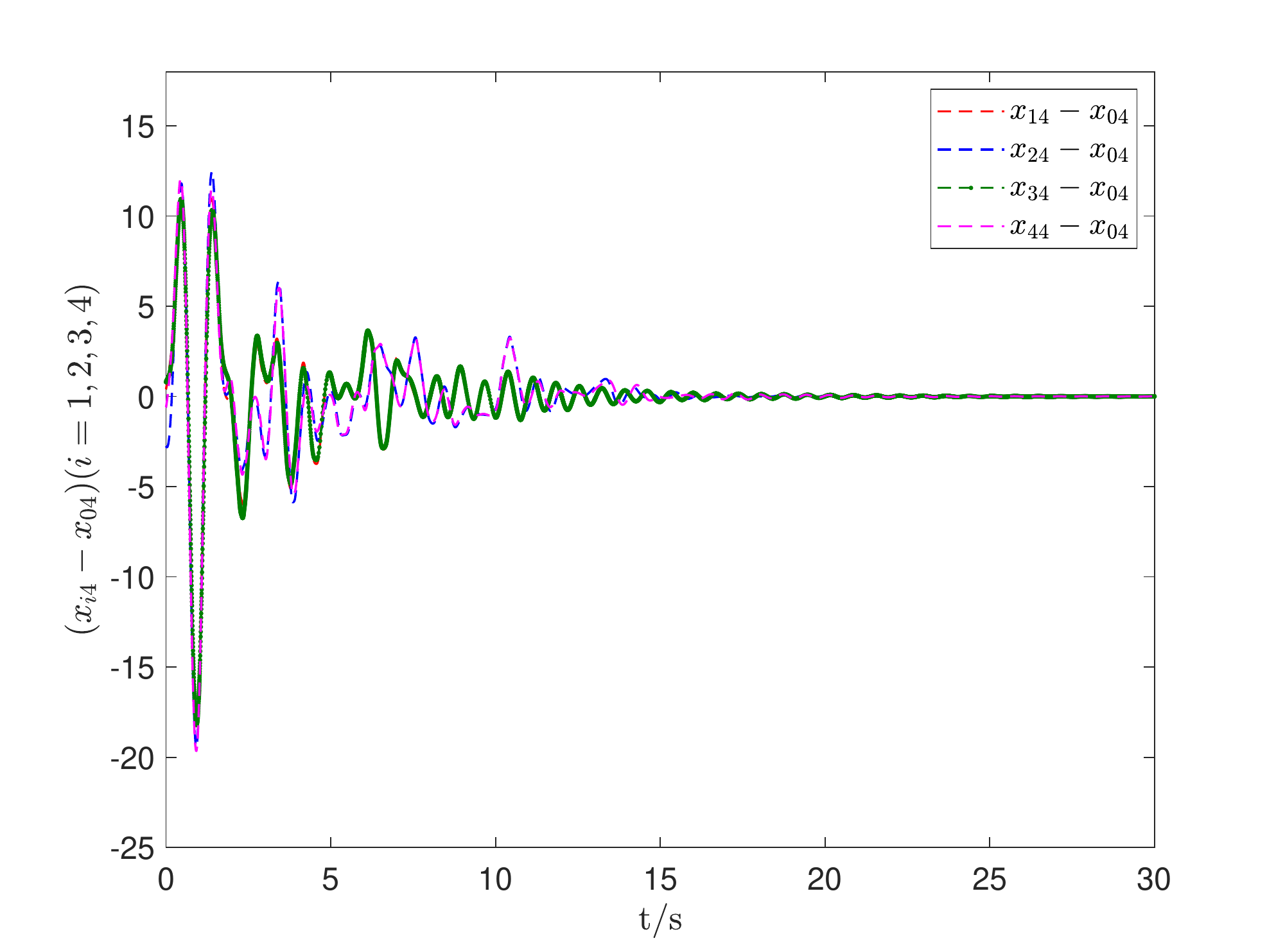}
		\caption{Tracking errors with $x_{i4}$ and $x_{04}$  under the designed controller (38)}
		\label{}
	\end{figure}

Figs. 12-15 demonstrate respectively the four states $x_{i1}, x_{i2}, x_{i3}$ and $x_{i4}$ of four agents could achieve the goal of consensus
control based on improved distributed coordinated tracking control protocol (38) with second intermittent coordinated constrained.
We can summarize that the proposed distributed
control protocol based on heterogeneous coupling framework and observers are effective and asymptotically stable for nonlinear MASs with intermittent communication and dynamic
switching topology.

	\section{Conclusions}
	
	In our work, the coordinated tracking control problem for nonlinear MASs under heterogeneous coupling network with intermittent communication has been studied.
Especially, considering the unavailable states, nonlinear observers have been constructed for each agents.
Then, a lemma has been correspondingly constructed to calculate the heterogeneous coupling gain, feedback gain and observer gain matrices. The system stability has been analyzed utilizing Lyapunov stability theory, switching system theory and LMI technology, and the allowed maximum communication threshold have been obtained. The results have been extended to the case of the intermittent communication with three types of intervals.
Finally, two simulation examples have been provided to prove the effectiveness and correctness of the proposed methods.

	\ifCLASSOPTIONcaptionsoff
	\newpage
	\fi
	

\end{document}